\documentclass[draftclsnofoot,onecolumn,12pt]{IEEEtran}
\usepackage{cite}
\usepackage{graphicx,color}
\usepackage{float}
\ifCLASSINFOpdf
\else
\fi
\usepackage[cmex10]{amsmath}
\DeclareMathOperator*{\argmax}{arg\,max}
\usepackage{bm}
\usepackage{amssymb}
\usepackage{algorithm}
\usepackage{algorithmic}
\usepackage{array}
\usepackage{fixltx2e}

\usepackage{indentfirst}
\usepackage{booktabs}
\usepackage{multirow}
\usepackage{bigstrut}
\usepackage{setspace}
\usepackage{booktabs}
\usepackage{xcolor}
\usepackage{amsthm}
\usepackage{epstopdf}
\usepackage{amsfonts}
\IEEEoverridecommandlockouts

\newtheorem{Theo}{Theorem}
\newtheorem{Lem}{Lemma}

\newtheorem{Def}{Definition}
\newtheorem{Rem}{Remark}

\newtheorem{Prop}{Proposition}

\begin{document}

\title{The Economics of Video Websites with Membership-Advertising Mode in Wireless Networks}
\author{\IEEEauthorblockN{Qing Zhou {\it{Student Member, IEEE}}, Nan Liu {\it{Member, IEEE}}}\\
}

\maketitle

\begin{abstract}
The popularization of smart mobile device and the rapid development of wireless communication technology lead to the explosion of global mobile data traffic, which brings huge business opportunities for content providers.
In this paper, we consider a novel business model of video websites via \textit{Membership-Advertising Mode} in wireless network, where the video websites provide three video services for mobile users: \textit{VIP-Member} service, \textit{Regular-Member} service and \textit{Non-Member} service.
The \textit{VIP-Member} (\textit{Regular-Member}) service provides the highest level (middle level) quality and non-advertising video service with high (low) price, while the \textit{Non-Member} service provides the lowest level quality and advertising-containing video service for free.
Meanwhile, the video websites sell their advertising spaces to the advertiser to create extra revenues.
We formulate the interactions among the advertiser, video websites and mobile users as a three-stage Stackelberg game.
Specifically, in Stage I, the advertiser decides the advertising budget; in Stage II, the video websites determine their advertising spaces selling strategies for advertiser and the membership pricing strategies for mobile users; in Stage III, the mobile users make their own decisions on video watching strategies for each video website.
We analyze the equilibrium of each sub-game.
Particularly, we derive the closed-form solutions of each mobile user's optimal video watching strategies, each video website's optimal membership price and the optimal advertising spaces selling number.
In addition, we also investigate the piece-wise structure of the advertiser's utility function, and further propose an efficient algorithm to obtain the optimal advertising budget.
Finally, numerical results show the impacts of different parameters' values on each entity's utility as well as the key indicators.
\end{abstract}

\IEEEpeerreviewmaketitle

\section{Introduction}
\subsection{Background}
With the spreading and application of smart mobile terminals, the global mobile data traffic grows explosively \cite{Tang2017}.
According to the prediction of Cisco, the global mobile data traffic is projected to achieve 49 exabytes/month by 2021, video streaming will occupy almost 78\% of the mobile traffic and the average mobile connection speed will exceed 20 Mbps\cite{Cisco2017}.
Meanwhile, the developments of recent communication technologies such as High-Density Network, Massive MIMO, Millimeter-Wave Communication and Local Caching, etc \cite{Boccardi2014,Agiwal2016,Shafi2017,Andrews2014} accommodate the requirements of data traffic explosion and data rate increment.

Under this context, the content providers are providing more and more of their services on the platform of the mobile device.
Take video websites for example, in the past, we have to watch live or recorded videos on PC by visiting video websites; nowadays, we can watch videos through the APPs on smart mobile devices.
Since the content providers (e.g., video websites) are profit-oriented entities, their objectives are to maximize their own revenues in addition to compensating for operational cost and the payment of video copyright.
Therefore, a new business model which is adapted to the mobile user-oriented network model is urgently needed for content providers to create more revenues \cite{Gabriel2014}.
\subsection{Motivations}
Recently, a variety of works have applied several game-theoretic approaches to study the business models of network resource allocation.
The literature can be categorized into several lines of works:
1) for cache resource allocation, references \cite{Liu2017,Li2016} respectively applied the Contract model and Stackelberg-game model to study the optimal pricing strategy of operators and the optimal caching strategy of content providers in a small-cell system, and reference \cite{Hajimirsadeghi2017} proposed the pricing and caching scheme in an information centric network.
2) For data offloading, reference \cite{Wang2016} utilized a distributed market framework to analyze the interactions between the offloading service providers (access points) and the offloading consumers (data traffic), reference \cite{Shah2017} proposed a three-stage Stackelberg game model to formulate the data offloading problem among mobile network operators and access point owners.
References \cite{Wang2009,Kang2012,Xu2016,Semasinghe2015} adopted different game approaches to investigate the physical parameters (e.g., power, bandwidth, SINR and average achievable rate) allocation problems, where \cite{Wang2009} studied the relay nodes selection and power allocation problems in cooperative relay model by the Stackelberg game approach, \cite{Kang2012} analyzed the price-based power allocation strategies for two-tier spectrum-sharing femtocell networks, \cite{Xu2016} investigated the wireless service provider selection problem and the corresponding bandwidth allocation problem in multi-tier heterogeneous cellular networks by evolutionary game and multi-leader multi-follower Stackelberg game, respectively; and \cite{Semasinghe2015} modeled multiple small cells by stochastic geometry and utilized evolutionary game theory to analyze the average SINR and the average achievable rate.
In addition, some works have also applied varieties of game theories to study the resource allocation problems in the fields of WIFI \& LTE-U \cite{Zhang2017,Gao2017,YuH2017,Yu2017}, distributed computing \cite{Yang2012Crowdsourcing,Duan2014}, cloud computing \& softwarized network\cite{Mansouri2017,Oro2017}, and rollover data \cite{Zheng2016,WangZ2017}, etc.
Although the above literatures have not directly considered the business model among content providers and mobile users, they have given us great insights for the business modeling of new scenarios and the analysis of the interactions among all the entities.

Consider the relationship among content providers and mobile users, and take video websites for example,
 \textit{Membership mode} is an effective business mode to create revenues \cite{Ribeiro2014}.
More specifically, Membership mode refers to an operational mode where the video website generates revenue from the mobile users through selling access rights of video resources to mobile users.

In addition to membership mode,
\textit{Advertising Mode}, an emerging business mode, provides another good solution for the content provider (e.g., video websites) monetization \cite{Flores2014,Bruner2000,Belch2004,Lavidge2000,Rossiter1987}.
In this mode, the video websites can create revenues from the advertisers via embedding the advertisements on their platforms in the form of text, picture or video.
Meanwhile, the advertisers can raise the brand awareness and further increase the sales of commodities via video advertising on video website, which is beneficial to increase the advertisers' revenue.
Several works have been conducted to study the field of advertising \cite{Bergemann2011,Johnson2013,Ghosh2015,Athey2016,Spentzouris2017,Nayak2017,Hosseinalipour2017,YuHao2017}.
More specifically, references \cite{Bergemann2011,Johnson2013,Ghosh2015,Athey2016} mainly focused on the study of the impact of targeting technology on advertising markets, references
\cite{Hosseinalipour2017,Nayak2017} analyzed the competitions among advertising companies in vehicular ad-hoc networks, reference
\cite{Spentzouris2017} investigated the advertising targeting and monetization problem of mobile location-based user in a venue (e.g., shopping mall or airport), and
\cite{YuHao2017} applied advertising to study the monetization model of the monopoly wifi business market.

Combining the above two business modes together, we can build a new mode called \textit{Membership-Advertising Mode}.
More specifically, the video website provides two types of services for mobile users: member service(s) (one or more) and non-member service.
The website members need to pay the video website for enjoying the high quality, ad-free video services, while the non-member can watch the low quality, ad-containing video services for free.
In this mode, the video website can create revenues from mobile user members as well as advertisers, and is thus a more flexible business mode than \textit{membership mode} or \textit{advertising Mode}.
In practice, some popular video websites in China, such as TENCNET, IQIYI, YOUKU and SOHU, have adopted the \textit{Membership-Advertising Mode} in their major business model.
However, we notice that few literature has studied the interactions among all interest entities of this model (e.g., advertiser, video websites and mobile users) as well as the designs of the corresponding policies.
Motivated by this, we conduct the study in this paper.

\subsection{Contributions}
In this paper, we investigate the economics of the video websites  with the \textit{Membership-Advertising Mode} in the wireless network.
More specifically, video websites provide three video services for mobile users: \textit{VIP-Member} service, \textit{Regular-Member} service and \textit{Non-Member} service.
For \textit{VIP-Member} and \textit{Regular-Member} services, the video websites charge the subscribed mobile users a fee according to their pricing mechanisms, where the price of the \textit{VIP-Member} service is higher than that of \textit{Regular-Member} service.
Correspondingly, \textit{VIP-Member} service subscribers can enjoy the high rate ad-free video service for unit period, and the \textit{Regular-Member} service subscribers can enjoy the middle rate ad-free video service for unit period. For \textit{Non-Member} services, although no membership fee is paid, the mobile users can only enjoy the low rate ad-containing video service.
According to the websites' pricing mechanisms, mobile users choose one of these three video services as their watching strategies based on their own preference of the video Quality-of-Experience (QoE).
Since the \textit{Non-Member} mobile users are the potential consumers of the advertiser, to promote the sales of its commodities, advertiser offers a certain budget for purchasing the advertising spaces among all video websites to display its commodities' advertisements.
Meanwhile, the advertising budget attracts each video website to sell its own advertising spaces to the advertiser.
Each video website decides the selling number of the advertising spaces with the consideration of the advertising budget, its own advertising space maintenance cost and other websites' advertising spaces selling strategies.

The main contributions of this paper can be summarized as follows:
\begin{itemize}
\item \textit{Novel Video Website's Business Model:} To the best of our knowledge, this is the first work that proposes and analyzes the network model consisting of multiple video websites, multiple mobile users and an advertiser.
  In this network model, we propose a novel \textit{Membership-Advertising Mode} in the monetization model.
\item \textit{Equilibrium Analysis:} We model the interactions among advertiser, video websites and mobile users as a three-stage Stackelberg game, and study the behaviour of each entity of interest as well as the corresponding sub-game equilibrium systematically. In particular, we derive the unique closed-form solutions of each mobile user's optimal watching strategies and each video website's optimal membership pricing strategy.
    We then determine the unique optimal subset of those video websites participating in the advertising budget game by exploiting the special structure of the advertisement (AD) space maintenance cost in the subset, and further derive the closed-form of the unique optimal AD-spaces selling number strategy for each website.
    Moreover, we analyze the piece-wise structure of the advertiser's utility and propose an efficient algorithm to obtain the optimal advertising budget.
\item \textit{Analysis of Key Parameters' impacts:} We analyze the relationship among the \textit{preference factor thresholds} with different values of  \textit{VIP-Member} price coefficient, and further investigate the coefficient's impact on mobile users' optimal watching strategies.
    Meanwhile, we also find that the subset of video websites that participate in the adverting budget game only depends on the AD-space maintenance cost rather than the advertising budget.
\item \textit{Performance Evaluation:} Extensive simulation results show that: 1) the increasing of \textit{VIP-Member} price coefficient is not beneficial for both video websites and advertiser to create revenues;
    2) The impact of \textit{popularity concentration level} on each utility is in connection with the popularity of the subset of the video websites that participate in the advertising budget game;
    3) The utility of video website (advertiser) increases (decreases) in the AD-space maintenance cost;
    4) The utility of video website (advertiser) increases in visiting frequency if the website's AD-spaces are \textit{saturated}, when visiting frequency is sufficiently large, the utility of video website (advertiser) is independent of visiting frequency.
\end{itemize}

\subsection{Organization}
The remainder of this paper is organized as follows. Section II introduces the system model and presents utility functions. Section III presents the three-stage stackelberg game to formulate the interactions of the utility function maximization problem and provides the game equilibrium. Section IV proposes the numerical computation to validate the results in this paper. Finally, section V concludes this paper.

\section{system model}
Before introducing the system model, we first list all the symbols in Table I for convenience of reading.
\begin{table}[htbp]\label{Symbol}
\caption{List of Symbols.}
\begin{tabular}{|l|l|}
\hline
\bf{Parameters} &\bf{Description}   \\ \hline
$K$ ~~~~~~&Video websites number. ~~~~\\ \hline
$N$ ~~~~~~&Mobile users number.~~~~\\  \hline
$G$ ~~~~~~&Advertisements number.~~~~\\  \hline
$Q_k^V$ ~~~~~~&\textit{VIP Member} service quality. ~~~~\\  \hline
$Q_k^R$ ~~~~~~&\textit{Regular Member} service quality. ~~~~\\  \hline
$Q_k^N$ ~~~~~~&\textit{Non Member}  service quality. ~~~~~\\  \hline
$\theta \in [0,\theta_{\max}]$ ~~~~~~&The \textit{preference} for QoE . ~~~~\\  \hline
$\sigma_k(>1)$ ~~~~~~&\textit{VIP Member} price coefficient.  ~~~~\\  \hline
$\gamma(>0)$~~~~~~&\textit{Concentration level}.~~~~\\  \hline
$\phi_k$ ~~~~~~& Video website $k$'s popularity.~~~~\\  \hline
$\xi (>0)$~~~~~~& Mobile user's visiting frequency. ~~~~\\  \hline
\end{tabular}
\begin{tabular}{|l|l|}
\hline
\bf{Variables} &\bf{Description}   \\ \hline
$C_k(>0)$~~~~~ ~~~~~~& AD space maitenance cost.~~~~~ ~~~~~\\  \hline
$p_k$ ~~~~~~~~~~~~&Membership price.\\  \hline
$M_k(P_a)$~~~~~~ ~~~~~&AD space selling number.\\  \hline
$P_a$ ~~~~~~~~~~~&The advertising budget.\\  \hline
\end{tabular}
\begin{tabular}{|l|l|}
\hline
\bf{Functions} &\bf{Description}   \\ \hline
${\chi}_k^V(p_k)$ & \textit{VIP Member} probability. \\  \hline
${\chi}_k^R(p_k)$ & \textit{Regular Member} probability. \\  \hline
${\chi}_k^N(p_k)$ & \textit{Non Member} probability. \\  \hline
$\bm{s}(\theta, \bm{p})$ &Mobile user's watching strategy set.  \\  \hline
$\tau_k(P_a, p_k,M_k)$ & \textit{AD Watching Probability}. \\  \hline
$V_k^{MU}(\theta,s_k,p_k)$&Mobile user's utility.\\  \hline
$V_k^{M}(p_k)$&  Video website's membership utility.\\  \hline
$V_k^{A}(\bm{M},P_a)$&  Video website's advertising utility.\\  \hline
$V_k^{VW}(\bm{M}, p_k,P_a)$&  Video website's total utility.\\  \hline
$V^{AD}(P_a, \boldsymbol{p},\boldsymbol{M})$ & The utility of the advertiser. \\  \hline
\end{tabular}
\end{table}

\begin{figure}[htbp]
\centering
\includegraphics[width=0.4\textwidth]{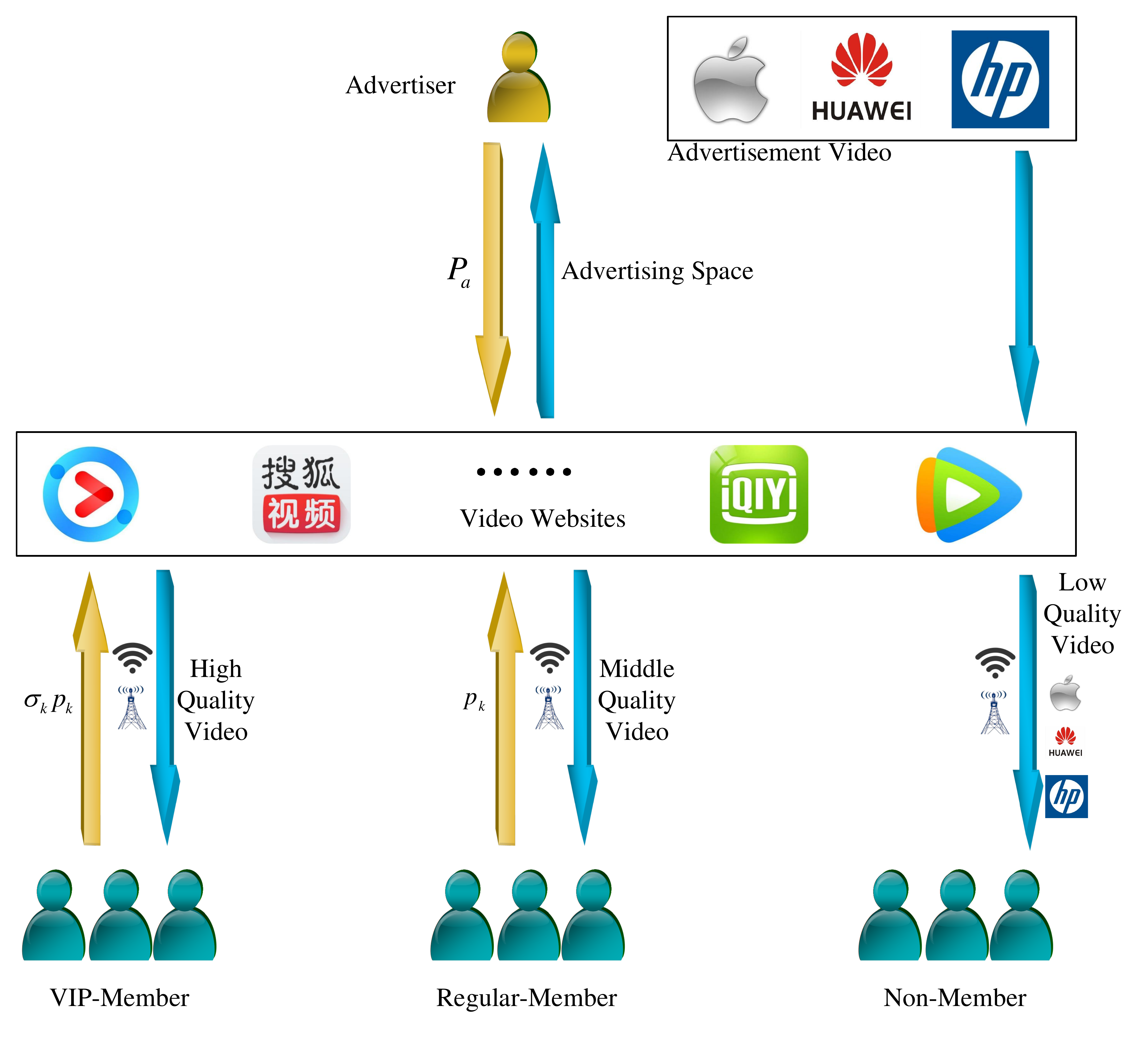}
\caption{System Model.}
\label{SystemModel}
\end{figure}

In this paper, we consider a network ecosystem which consists of an advertiser, $K$ video websites and $N$ mobile users, as show in Fig. \ref{SystemModel}.
We focus on the interactions among them via a three-stage Stackelberg game.

\subsection{Network Model}
\subsubsection{Advertiser}
The advertiser has total $G$ advertisements (ADs) which seek to display on $K$ video websites.
The advertiser has total $P_a$ advertising budget to buy the advertisement (AD) spaces from $K$ video websites.
Here, the unit AD-space refers to the time segment (e.g., 30 seconds) at the beginning of a video which is specially used for displaying video AD.

\subsubsection{Video Website}
There are total $K$ video websites in the network, and we denote $\mathcal{K}\triangleq \{1,2,\cdots,K\}$ as the website set.
There exists two-fold interactions for each video website: one is the interaction between the video website and the advertiser: under the motivation of the advertising budget $P_a$, all video websites compete with each other through selling the AD spaces $M_k$; the other is the interaction between the video website and the mobile users: each video website $k$ provides three video services for mobile users: \textit{VIP-Member} service, \textit{Regular-Member} service and \textit{Non-Member} service.
We let $Q_k^V$, $Q_k^R$ and $Q_k^N$ respectively denote the corresponding video service quality of website $k$.
More specifically,  for the \textit{VIP-Member} service, the website $k$ charges its subscribed mobile user $\sigma_kp_k$ ($\sigma_k>1$), where $\sigma_k$ is the \textit{VIP-Member} price coefficient, and provides the video service with video rate $r_k^V$ per unit period (e.g., /month), and the service quality is measured by $Q_k^V= \text{ln}(1+r_k^V)$ \cite{JianweiHuang2014};
for the \textit{Regular-Member} service, the website $k$ charges its subscribed mobile user $p_k$, and provides the non-advertising video service with video rate $r_k^R$  per unit period, and the service quality is measured by $Q_k^R= \text{ln}(1+r_k^R)$ \cite{JianweiHuang2014};
for the \textit{Non-Member} service, although it is free of charge, the webwite only provides the advertising-containing video service with video rate $r_k^N$, and the service quality is measured by $Q_k^N= \text{ln}(1+r_k^N)$ \cite{JianweiHuang2014}. Without loss of generality, we assume that $r_k^V > r_k^R > r_k^N >0$, naturally, we have $Q_k^V > Q_k^R > Q_k^N >0$.

\subsubsection{Mobile User}
We denote $N$ as the total number of mobile users browsing the video websites within an unit period.
We let a positive $\theta$ denote each user's \textit{preference} for Quality-of-Experience (QoE) for website browsing, and further assume $\theta$ follows the uniform distribution over $[0,\theta_{\max}]$.
The \textit{preference} factor $\theta$ reflects the user's requirement and willingness for video watching experience, namely, a larger $\theta$ implies that the user is sensitive to the video QoE, while a lower $\theta$ means the user is relatively QoE-tolerant.
Note that since different users may have different \textit{preference} factor $\theta$ values, they have different watching strategies when browsing one video website $k$, $k\in \mathcal{K}$.

According to the video service types of each video website, we denote $s_k \in \{1,2,3\}$ as a mobile user's corresponding watching strategy when browsing video website $k$, and denote $\boldsymbol{s}\triangleq\{s_1, s_2, \cdots, s_K\}$ as the mobile user's watching strategy set for all the $K$ websites.
Specifically, $s_k = 1$, $s_k = 2$ and $s_k = 3$ denote that an mobile user chooses to be an \textit{VIP-Member}, \textit{Regular-Member} and \textit{Non-Member} of website $k$, respectively.

\subsection{Mobile User's Utility}
\subsubsection{Mobile User's Payoff}
Based on mobile user's three different watching strategies, for a type-$\theta$ user browsing website $k$, the payoff over an unit period is defined as follows:
\begin{align}\label{Payoff}
V_k^{MU}(\theta,s_k,p_k) =
\left\{ {\begin{array}{lr}
\theta Q_k^V-\sigma_kp_k, &{s_k = 1;}\\
\theta Q_k^R-p_k, &{s_k = 2;}\\
\theta Q_k^N, &{s_k = 3.}
\end{array}} \right.
\end{align}

Each mobile user will choose one strategy which maximizes the payoff as the utility function.

\subsubsection{Strategy Probability, Visiting Frequency and Websites Preference}
Given price $p_k$, we denote ${\chi}_k^V(p_k)$, ${\chi}_k^R(p_k)$ and ${\chi}_k^N(p_k)$ as the probability that a mobile user chooses to be \textit{VIP-Member}, \textit{Regular-Member} and \textit{Non-Member}, respectively.
In addition, we assume that each mobile user's visiting frequency for each video website per unit period follows the Poisson distribution with mean $\xi (>0)$.
Meanwhile, due to different influence factors such as individual willingness, QoS, charge standard, video content popularity and so on, mobile user has different level of preference for all the $K$ video website.
We let $\boldsymbol{\phi} \triangleq [\phi_1, \phi_1, \cdots, \phi_K]$ denote each mobile user's preference of $K$ video websites, where $\phi_k$ follows the Zipf distribution, i.e.,
\begin{align}
\phi_k = \frac{1/k^\gamma}{\sum_{i=1}^K {\frac{1} { i^\gamma}}}
\end{align}
for $k\in \mathcal{K}$.
Note that this assumption is also made in \cite{Li2016,Liu2017}.
The parameter $\gamma$  is a positive real number which characterizes the \textit{concentration level}  of the website's popularity distribution.
A larger $\gamma$ implies that the polarization of popularity is more obvious, that is, the popularity of those websites with smaller index $k$ is much higher than others.
\subsection{Video Website's Utility}
Since each video website has the interactions with both mobile users and advertiser, its utility comes from two aspects.

One is the utility comes from the mobile users' payments for membership and we denote it by $V_k^M(p_k)$, $k\in \mathcal{K}$:
\begin{align}\label{Utility_Member}
   V_k^M(p_k) =  \sigma_kp_k{\phi_k N{\chi}_k^V(p_k)}+p_k{\phi_k N{\chi}_k^R(p_k)}
\end{align}
Recall that $N$ is the mobile user number, $\phi_k$ is the popularity of video website $k$, ${\chi}_k^V(p_k)$ and ${\chi}_k^R(p_k)$ are respectively the \textit{VIP-Member} probability and the \textit{Regular-Member} probability of website $k$ under price $p_k$, thus $\phi_k N{\chi}_k^V(p_k)$ and $\phi_k N{\chi}_k^R(p_k)$ are respectively the expected number of the \textit{VIP-Member} and the \textit{Regular-Member} of website $k$.

The other is utility comes from selling AD spaces to the advertiser. We assume the reward that video website $k$ receives is proportional to $M_k$, and denote $C_k$ as the maintenance cost of unit AD space, then, the utility of website $k$ in terms of selling AD space can be expressed as follows:
\begin{align}\label{AD_Utility}
V_k^A(M_k,\bm{M}_{-k},P_a)  = {\frac {M_k}{ \sum_{j=1}^K {M_j}}}P_a  - C_kM_k, k\in \mathcal{K}
\end{align}
where $\bm{M}_{-k} = \{M_i|i \in \mathcal{K}\backslash k\}$.

To sum up, the utility of video website $k$, $k\in \mathcal{K}$, $V_k^{VW}(M_k,\bm{M}_{-k}, p_k,P_a) $ is given by:
\begin{align}
V_k^{VW}(M_k,\bm{M}_{-k}, p_k,P_a) =  V_k^M(p_k) + V_k^A(M_k,\bm{M}_{-k},P_a)
\end{align}

\subsection{Advertiser's Utility}
\subsubsection{AD Watching Probability}
We assume the total number of AD spaces that video website $k$ sells to the advertiser under advertising price $P_a$ is $M_k(P_a)$.
Meanwhile, we assume that the AD spaces allocation among all the ADs follow \textit{equal allocation principle}.
Since each website needs to display total $G$ ADs, the total display number of each AD over unit period is $\frac{M_k(P_a)}{G}$.
In addition, the expected number of mobile users who watch the advertising-containing video of website $k$ is $\phi_k N{\chi}_k^N(p_k)$.
Since the average visiting frequency of each mobile user per unit period is $\xi$, the total visiting number of the advertising-containing video of website $k$ per unit period is $\xi\phi_k N{\chi}_k^N(p_k)$.
Therefore, if a mobile user watches an advertising-containing video of website $k$, the probability that the user can watch a specific advertisement is given by:
\begin{align}\label{ADWP}
\varphi_k(P_a, p_k,M_k) = {\frac{M_k(P_a)} {G\xi\phi_k N{\chi}_k^N(p_k)}}, k \in \mathcal{K}.
\end{align}
Note that each mobile user's website visiting frequency is a random discrete variable $X$ which follows Poisson distribution with mean $\xi$, and its probability mass function (PMF) is given by \cite{Gallager2013}:
\begin{align}
\mathbb{P}[X] = e^{-\xi}{\frac{\xi^{X}}{X!}}, X = 0, 1, \cdots, \infty.
\end{align}
According to the complementary advertising assumption \cite{Bagwell2007,Bergemann2011,Guo2017}, we know that the advertiser can receive the reward of an advertisement on video website $k$  only if the specific advertisement has been watched by the mobile user at least one time.
We further define the probability that a specific advertisement on video website $k$ which has been watched at least one time as \textit{AD Watching Probability}, i.e., $\tau_k(P_a, p_k,M_k)$, which is given by:
\begin{align}\label{ADWatchPro}
\tau_k(P_a, p_k,M_k) &= 1 -\sum_{x = 0}^{\infty}\mathbb{P}[x](1-\varphi_k(P_a, p_k,M_k))^x
                                = 1 - \sum_{x = 0}^{\infty}e^{-\xi}{\frac{\xi^{x}}{x!}}\left(1-{\frac{M_k(P_a)} {G\xi\phi_k N{\chi}_k^N(p_k)}}\right)^x \nonumber\\
                                & = 1- e^{-\xi} \left[ \sum_{x = 0}^{\infty} \frac{\left(\xi-{\frac{M_k(P_a)}{G\phi_k N{\chi}_k^N(p_k)}}\right)^x}{{x!}}\right]
                                \overset{(a)}{=} 1 - \exp\left(-\frac{M_k(P_a)}{G\phi_k N{\chi}_k^N(p_k)}\right)
\end{align}
where (a) is obtained by the Taylor expansion for exponential function $e^x = \sum_{i=0}^{\infty} {\frac{x^i}{i!}}$.

\subsubsection{Advertiser's Payoff}
Based on the \textit{AD Watching Probability} $\tau_k(P_a, p_k,M_k)$, $k \in \mathcal{K}$, the payoff of the advertiser $V^{AD}(P_a, \boldsymbol{p},\boldsymbol{M})$ is given by:
\begin{align}\label{ADer_Utility}
V^{AD}(P_a, \boldsymbol{p},\boldsymbol{M}) = \sum_{k=1}^K \omega_k G \phi_k N {\chi}_k^N(p_k) \tau_k(P_a, p_k,M_k) - P_a
\end{align}
where $\omega_k$ is the reward coefficient that the advertiser display ADs on website $k$, $G$ is the total ADs number, $\phi_k N {\chi}_k^N(p_k)$ is the expected \textit{Non-Member} number of website $k$, $\tau_k(P_a, p_k,M_k) $ is the \textit{AD Watching Probability}, $P_a$ is the advertising budget, $\boldsymbol{p} \triangleq \{p_1,p_2,\cdots,p_K\}$ is the set of \textit{Regular-Member} price and $\boldsymbol{M} \triangleq \{M_1,M_2,\cdots,M_K\}$ is the set of selling number of AD spaces.

\section{Three-Stage Stackelberg Game}
In this section, we will utilize a three-stage stackelberg game to formulate the interactions among the advertiser, video websites and the mobile users.
Specifically, in Stage I, the advertiser decides its total advertising budget $P_a$;
in Stage II, each video website $k$ specifies the selling number of AD spaces $M_k$ and the \textit{Regular-Member} price $p_k$;
in Stage III, each type-$\theta$ mobile user chooses the video website watching strategies $\boldsymbol{s}$.
We use backward induction to analyze the three-stage Stackelberg game.

\subsection{Stage III: Mobile User's Optimal Video Watching Strategy $\boldsymbol{s}^*$}
\subsubsection{{Preference} Factor Threshold}
For a specific video website $k$, the goal of each mobile user is to choose one video watching strategy $s_k$ that maximizes its payoff in (\ref{Payoff}).
Since the \textit{Regular Member} price $p_k$ in (\ref{Payoff}) is given in Stage II, each mobile user's video watching strategy $\boldsymbol{s}$ only depends on the \textit{preference} factor $\theta \in [0, \theta_{\max}]$.
Before we proceed, we first give some definitions about the \textit{preference factor threshold}.
\begin{Def}\label{Threshold1}
\rm{
We define $\theta_k^{T_{12}}$ as the \textit{preference factor threshold} such that $V_k^{MU}(\theta,s_k=1,p_k) = V_k^{MU}(\theta,s_k=2,p_k)$, then, $\theta_k^{T_{12}} \triangleq \min\{ {\frac{(\sigma_k-1)p_k}{Q_k^V- Q_k^R}}, \theta_{\max}\}$.
}
\end{Def}

\begin{Def}\label{Threshold2}
\rm{
We define $\theta_k^{T_{13}}$ as the \textit{preference factor threshold} such that $V_k^{MU}(\theta,s_k=1,p_k) = V_k^{MU}(\theta,s_k=3,p_k)$, then, $\theta_k^{T_{13}} \triangleq \min\{{\frac{\sigma_kp_k}{Q_k^V- Q_k^N}}, \theta_{\max}\}$.
}
\end{Def}

\begin{Def}\label{Threshold3}
\rm{
We define $\theta_k^{T_{23}}$ as the \textit{{preference} factor threshold} such that $V_k^{MU}(\theta,s_k=2,p_k) = V_k^{MU}(\theta,s_k=3,p_k)$, then, $\theta_k^{T_{23}} \triangleq \min\{ {\frac{p_k}{Q_k^R- Q_k^N}}, \theta_{\max}\}$.
}
\end{Def}

Based on the \textit{{preference} factor threshold} of video website $k$ ($\in \mathcal{K}$), the optimal video watching strategy of a type-$\theta$ mobile user is given by:
\begin{align}\label{Watching_Strategy}
s_k^*(\theta, p_k) =
\left\{ {\begin{array}{rr}
1, &\text{if}~\max\{\theta_k^{T_{12}}, \theta_k^{T_{13}}\} \le \theta \le \theta_{\max}; \\
2, &\text{if}~\theta_k^{T_{23}} \le \theta \le \theta_k^{T_{12}};\\
3, &\text{if}~0\le \theta \le \min\{\theta_k^{T_{23}}, \theta_k^{T_{13}}\}.
\end{array}} \right.
\end{align}

\subsubsection{The impact of  $\sigma_k$ on $s_k^*$}
\begin{figure}[htb]
\centering
\includegraphics[width=0.4\textwidth]{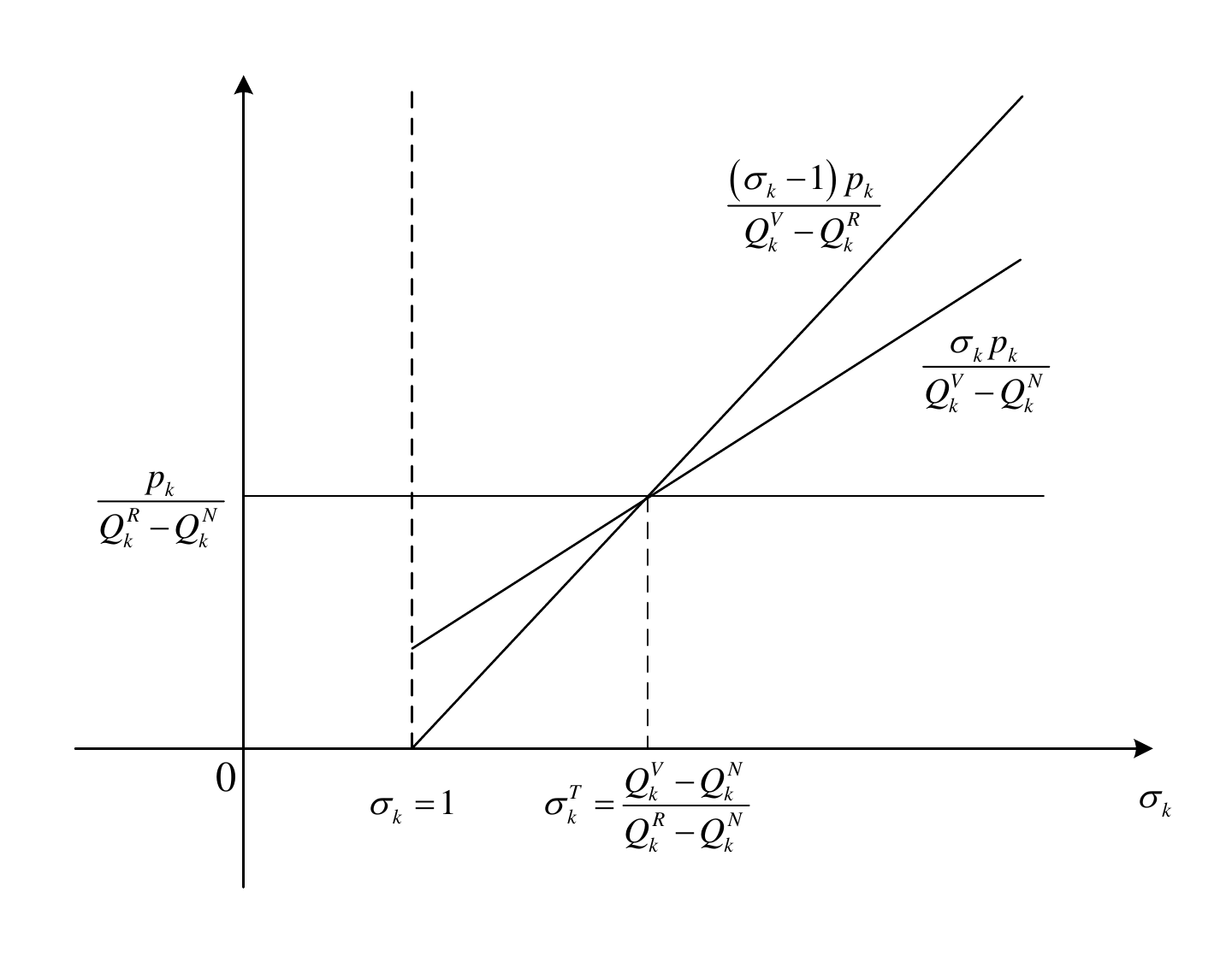}
\caption{Graphical Visualization of Lemma \ref{Threshold_Relationship}.}
\label{Graphical Visualization}
\end{figure}

Next, we will analyze the relationship among ${\frac{(\sigma_k-1)p_k}{Q_k^V- Q_k^R}}$, ${\frac{\sigma_kp_k}{Q_k^V- Q_k^N}}$ and ${\frac{p_k}{Q_k^R- Q_k^N}}$, and rigorously describe it in the following lemma.
\begin{Lem} \label{Threshold_Relationship}
\rm
{
For $k \in \mathcal{K}$,
\begin{itemize}
\item[1.] when $1<\sigma_k \le \sigma_k^T \triangleq {\frac{Q_k^V-Q_k^N}{Q_k^R-Q_k^N}}$, then,
${\frac{(\sigma_k-1)p_k}{Q_k^V- Q_k^R}} \le {\frac{\sigma_kp_k}{Q_k^V- Q_k^N}} \le {\frac{p_k}{Q_k^R- Q_k^N}}$;
\item[2.] when $\sigma_k > \sigma_k^T \triangleq {\frac{Q_k^V-Q_k^N}{Q_k^R-Q_k^N}}$, then,
${\frac{(\sigma_k-1)p_k}{Q_k^V- Q_k^R}} > {\frac{\sigma_kp_k}{Q_k^V- Q_k^N}} > {\frac{p_k}{Q_k^R- Q_k^N}}$.
\end{itemize}
}
\end{Lem}
\begin{IEEEproof}
Please refer to Appendix \ref{Proof_Threshold_Relationship}.
\end{IEEEproof}
\begin{Rem}
\rm{
Firstly, it is obvious that both ${\frac{(\sigma_k-1)p_k}{Q_k^V- Q_k^R}}$ and ${\frac{\sigma_kp_k}{Q_k^V- Q_k^N}}$ are the linear increasing function of $\sigma_k$ since both ${\frac{p_k}{Q_k^V- Q_k^R}}$ and ${\frac{p_k}{Q_k^V- Q_k^N}}$  are two positive constants under price $p_k$.
Moreover, since ${\frac{p_k}{Q_k^V- Q_k^R}}>{\frac{p_k}{Q_k^V- Q_k^N}}>0$, the slope of straight line ${\frac{(\sigma_k-1)p_k}{Q_k^V- Q_k^R}}$, i.e., ${\frac{p_k}{Q_k^V- Q_k^R}}$, is larger than that of straight line ${\frac{\sigma_kp_k}{Q_k^V- Q_k^N}}$, i.e., ${\frac{p_k}{Q_k^V- Q_k^N}}$;
Secondly, ${\frac{p_k}{Q_k^R- Q_k^N}}$ is a $\sigma_k$-independent positive constant;
Thirdly, since these three straight lines intersect at $\sigma_k = \sigma_k^T \triangleq {\frac{Q_k^V-Q_k^N}{Q_k^R-Q_k^N}}$, in the domain $(1, \sigma_k^T]$, we have ${\frac{(\sigma_k-1)p_k}{Q_k^V- Q_k^R}} \le {\frac{\sigma_kp_k}{Q_k^V- Q_k^N}} \le {\frac{p_k}{Q_k^R- Q_k^N}}$, and in the domain $(\sigma_k^T, \infty)$, we have ${\frac{(\sigma_k-1)p_k}{Q_k^V- Q_k^R}} > {\frac{\sigma_kp_k}{Q_k^V- Q_k^N}} > {\frac{p_k}{Q_k^R- Q_k^N}}$.
Here, we present the graphical visualization of the relationship among ${\frac{(\sigma_k-1)p_k}{Q_k^V- Q_k^R}}$, ${\frac{\sigma_kp_k}{Q_k^V- Q_k^N}}$ and ${\frac{p_k}{Q_k^R- Q_k^N}}$ stated in Lemma \ref{Threshold_Relationship} in Fig. \ref{Graphical Visualization}.
}
\end{Rem}

When taking $\theta_{\max}$ into consideration, there are total five possibilities of the optimal video watching strategy $s_k^*$ to discuss.

{\bf{Case 1--$\sigma_k \in (1, \sigma_k^T]$}:} Combining Definition \ref{Threshold1}-\ref{Threshold3} and Lemma \ref{Threshold_Relationship}, it is not difficult to find out that the inequality $\theta_k^{T_{12}} \le \theta_k^{T_{13}} \le \theta_k^{T_{23}}$ holds for any positive $\theta_{\max}$\footnote{According to Definition \ref{Threshold1}-\ref{Threshold3},  when $\theta_{\max} \in [0, \theta_k^{T_{12}}]$, we have $\theta_k^{T_{12}} = \theta_k^{T_{13}}= \theta_k^{T_{23}}= \theta_{\max}$;
when $\theta_{\max} \in ( \theta_k^{T_{12}}, \theta_k^{T_{13}}]$, we have $\theta_k^{T_{12}} = {\frac{(\sigma_k-1)p_k}{Q_k^V- Q_k^R}}$, $\theta_k^{T_{13}} =\theta_k^{T_{23}}= \theta_{\max}$, thus $\theta_k^{T_{12}} \le \theta_k^{T_{13}} = \theta_k^{T_{23}}$;
when $\theta_{\max} \in ( \theta_k^{T_{13}}, \theta_k^{T_{23}}]$, we have $\theta_k^{T_{12}} = {\frac{(\sigma_k-1)p_k}{Q_k^V- Q_k^R}}$, $\theta_k^{T_{13}} = {\frac{\sigma_kp_k}{Q_k^V- Q_k^N}}$, $\theta_k^{T_{13}} = \theta_{\max}$, thus $\theta_k^{T_{12}} \le \theta_k^{T_{13}} \le \theta_k^{T_{23}}$;
when $\theta_{\max} \in (\theta_k^{T_{23}}, \infty)$, we have $\theta_k^{T_{12}} = {\frac{(\sigma_k-1)p_k}{Q_k^V- Q_k^R}}\le \theta_k^{T_{13}} = {\frac{\sigma_kp_k}{Q_k^V- Q_k^N}} \le \theta_k^{T_{23}}= {\frac{p_k}{Q_k^R- Q_k^N}}$.}.
Therefore, the optimal video watching strategy of a type-$\theta$ mobile user in (\ref{Watching_Strategy}) can be reduced to:
\begin{align}\label{Watching_Strategy_New}
s_k^*(\theta, p_k) =
\left\{ {\begin{array}{lr}
1, &\text{if}~\theta_k^{T_{13}} \le \theta \le \theta_{\max}; \\
3, &\text{if}~0\le \theta \le \theta_k^{T_{13}}.
\end{array}} \right. k\in \mathcal{K}.
\end{align}

\begin{figure}[htb]
\centering
\includegraphics[width=0.4\textwidth]{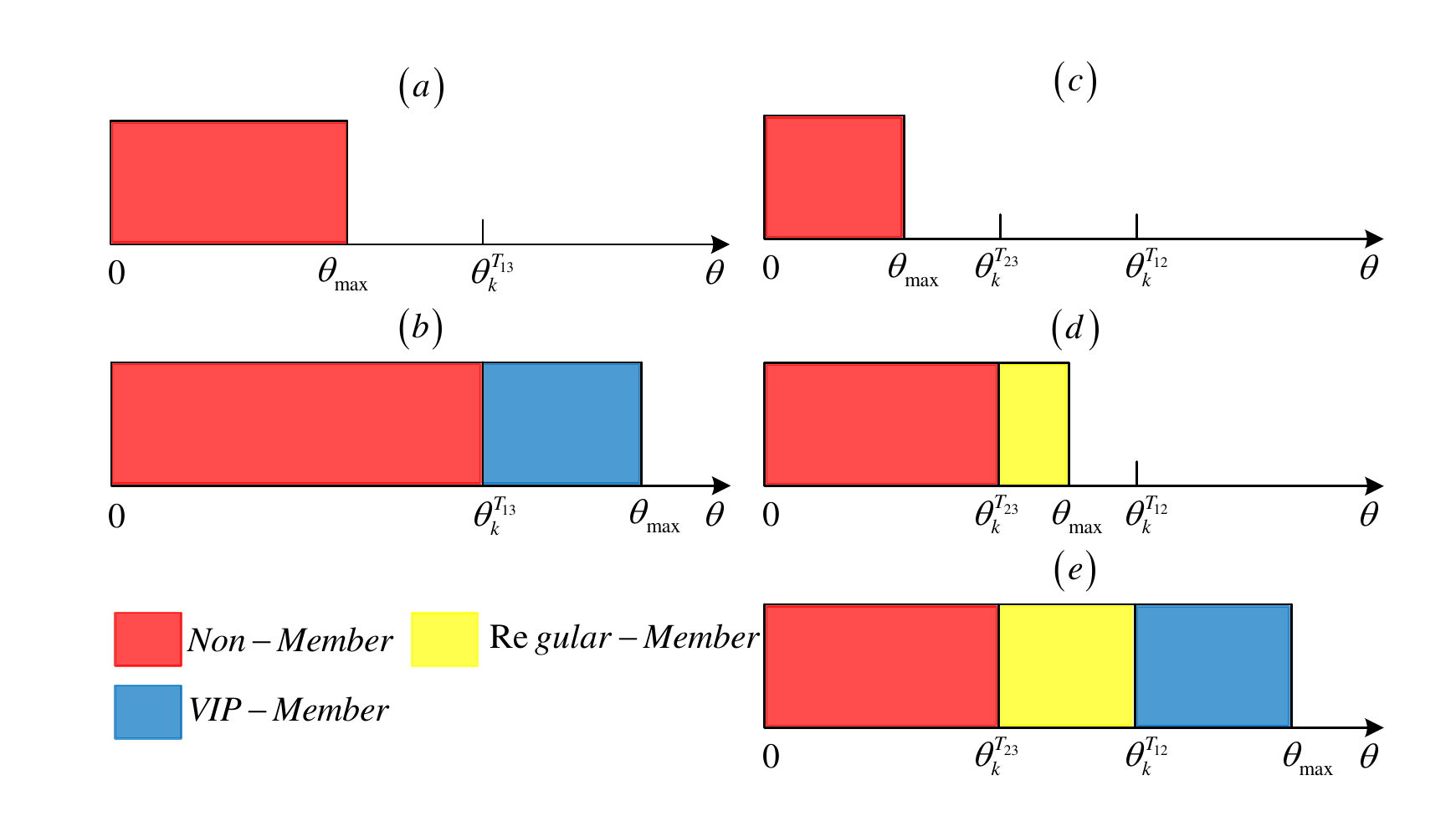}
\caption{Graphical illustration of all the optimal video watching strategies.
 (a) and (b) show the two optimal strategies in the domain $\sigma_k \in (1, \sigma_k^T]$: in (a), $\theta_{\max}\in [0,\theta_k^{T_{13}}]$, $s_k^* = 3$ for $\theta \in [0,\theta_{\max}]$; in (b), $\theta_{\max}\in (\theta_k^{T_{13}}, \infty)$, $s_k^* = 3$ for $\theta \in [0,\theta_k^{T_{13}}]$ and $s_k^* = 1$ for $\theta \in (\theta_k^{T_{13}},\theta_{\max}]$.
  (c), (d) and (e) show the three optimal strategies in the domain $\sigma_k \in (\sigma_k^T, \infty)$: in (c), $\theta_{\max}\in [0,\theta_k^{T_{23}}]$, $s_k^* = 3$ for $\theta \in [0,\theta_{\max}]$; in (d), $\theta_{\max}\in (\theta_k^{T_{23}}, \theta_k^{T_{12}}]$, $s_k^* = 3$ for $\theta \in [0,\theta_k^{T_{23}}]$ and $s_k^* = 2$ for $\theta \in (\theta_k^{T_{23}},\theta_{\max}]$;
  in (e), $\theta_{\max}\in (\theta_k^{T_{12}}, \infty)$, $s_k^* = 3$ for $\theta \in [0,\theta_k^{T_{23}}]$, $s_k^* = 2$ for $\theta \in (\theta_k^{T_{23}},\theta_k^{T_{12}}]$ and $s_k^* = 1$ for $\theta \in (\theta_k^{T_{12}},\theta_{\max}]$.}.
\label{Watching Strategy}
\end{figure}

{\bf{Case 2--$\sigma_k \in (\sigma_k^T,\infty)$}:} Similar to the analysis of the footnote in case 1, the inequality $\theta_k^{T_{12}} \ge \theta_k^{T_{13}} \ge \theta_k^{T_{23}}$ holds for any positive $\theta_{\max}$.
Therefore, the corresponding optimal video watching strategy in (\ref{Watching_Strategy}) can be further reduced to:
\begin{align}\label{Watching_Strategy_New1}
s_k^*(\theta, p_k) =
\left\{ {\begin{array}{rr}
1, &\text{if}~\theta_k^{T_{12}} \le \theta \le \theta_{\max}; \\
2, &\text{if}~\theta_k^{T_{23}} \le \theta \le \theta_k^{T_{12}};\\
3, &\text{if}~0\le \theta \le \theta_k^{T_{23}}.
\end{array}} \right. k\in \mathcal{K}.
\end{align}
In addition, The graphical illustration of all the five possibilities of the optimal video watching strategy $s_k^*$ stated in (\ref{Watching_Strategy_New}) and (\ref{Watching_Strategy_New1}) is presented in Fig. \ref{Watching Strategy}.

In order to guarantee that the feasible domain of each video watching strategy (i.e., $s_k = 1, 2, 3$) is non-empty, without loss of generality, we only consider the case such that $\sigma_k \in (\sigma_k^T,\infty)$ and $\theta_{\max}\in (\theta_k^{T_{12}}, \infty)$, as shown in Fig. \ref{Watching Strategy}(e).
The analysis of case in Fig. \ref{Watching Strategy}(e) can be applied in all other cases.
\subsubsection{Watching Strategy Probability}
Note that $\theta$ follows the uniform distribution over $[0, \theta_{\max}]$, therefore, under $p_k$ and coefficient $\sigma_k$, the probability that a mobile user chooses to be \textit{VIP-Member}, \textit{Regular-Member} and \textit{Non-Member} of website $k$ can be respectively given by:
\begin{align}
{\chi}_k^V(p_k) &= {\frac{\theta_{\max}-\theta_k^{T_{12}}}{\theta_{\max}}} = 1-{\frac{(\sigma_k-1)p_k}{(Q_k^V- Q_k^R)\theta_{\max}}}.\label{Prob_VIP}\\
{\chi}_k^R(p_k) &= {\frac{\theta_k^{T_{12}}- \theta_k^{T_{23}}}{\theta_{\max}}}
={\frac{p_k\left[\sigma_k( Q_k^R- Q_k^N)- ( Q_k^V- Q_k^N)\right]}{( Q_k^V- Q_k^R)( Q_k^R- Q_k^N)\theta_{\max}}}.\label{Prob_Reg}\\
{\chi}_k^N(p_k) &= {\frac{\theta_k^{T_{23}}}{\theta_{\max}}} = {\frac{p_k}{( Q_k^R- Q_k^N)\theta_{\max}}}.\label{Prob_Non}
\end{align}

\subsection{Stage II: Video Websites' Optimal Membership Pricing  $\boldsymbol{p}^* $ and Optimal Selling Number of AD Spaces $\boldsymbol{M}^*$}
In this subsection, we consider the video Websites' optimal membership pricing strategies $\boldsymbol{p}^* \triangleq \{p_1^*,p_2^*,\cdots,p_K^*\}$ for mobile users and the optimal selling number of AD spaces $\boldsymbol{M}^* \triangleq \{M_1^*,M_2^*,\cdots,M_K^*\}$ for advertiser in stage II.
Each video website $k (\in \mathcal{K})$ decides its optimal member pricing $p_k^*$ by anticipating the mobile user's watching strategy in stage III, and makes its decision on optimal selling number of AD space $M_k^*$ not only by responding to the advertising budget in stage I, but also depending on the AD spaces selling strategies of other websites $\bm{M}_{-k}^*$.

\subsubsection{Optimal membership pricing $p_k^*$}
Substituting (\ref{Prob_VIP}) and (\ref{Prob_Reg}) into (\ref{Utility_Member}), the optimization problem of maximizing the utility of video website $k$ in terms of mobile users' payments for membership can be formulated as:
\begin{align}
\max_{p_k} V_k^M(p_k) &=  \max_{p_k} \phi_k N\left(\sigma_kp_k{\chi}_k^V(p_k)+p_k{\chi}_k^R(p_k)\right)\nonumber\\
&= \max_{p_k} \phi_k N \left[\sigma_k p_k - {p_k^2{\frac {(\sigma_k-1)^2( Q_k^R- Q_k^N)+( Q_k^V- Q_k^R)}{\theta_{\max}( Q_k^V- Q_k^R)( Q_k^R- Q_k^N)}}}\right].\label{Problom_utility_webprice}\\
&\text{s.t.}~\left\{ {\begin{array}{lr}\label{Problom_utility_webprice Contraint}
&p_k \ge 0.\\
&\theta_k^{T_{12}}(p_k) \le \theta_{\max}.
\end{array}} \right. k\in \mathcal{K}.
\end{align}
where the first constraint in (\ref{Problom_utility_webprice Contraint}) means the membership price is non-negative, and the second constraint in (\ref{Problom_utility_webprice Contraint}) means the \textit{preference factor threshold} $\theta_k^{T_{12}}(p_k)$ should no larger than $\theta_{\max}$ to guarantee non-empty feasible domain of each mobile user's video watching strategy (i.e., $s_k = 1, 2, 3$).
\begin{Theo}\label{Optimal web price}
\rm
{
The unique optimal membership price of video website $k$, i.e., $p_k^*$, for problem in (\ref{Problom_utility_webprice}) is given by:
\begin{align}\label{Opt_MemPrice}
p_k^* = \left[ {\frac{\sigma_k\theta_{\max}( Q_k^V- Q_k^R)( Q_k^R- Q_k^N)}{2\left[ (\sigma_k-1)^2( Q_k^R- Q_k^N)+( Q_k^V- Q_k^R)\right]}}\right]^{\frac{\theta_{\max}( Q_k^V- Q_k^R)}{(\sigma_k-1)}}
\end{align}
where $[a]^b \triangleq \min\{a, b\}$.
}
\end{Theo}
\begin{IEEEproof}
Please refer to Appendix \ref{Proof_Optimal web price}.
\end{IEEEproof}
\subsubsection{Optimal Selling Number of AD Spaces $\boldsymbol{M}^*$}
In this part, we will determine the optimal selling number of AD spaces $\boldsymbol{M}^* \triangleq \{M_1^*,M_2^*,\cdots,M_K^*\}$.
From the utility function of each video website $k$ in (\ref{AD_Utility}), we can see that the optimal selling number of the AD spaces of website $k$, i.e., $M_k^*$, not only depends on the total advertising budget $P_a$, but also depends on the AD spaces selling strategies of other websites, i.e., $\boldsymbol{M}_{-k}^*$. Therefore, the process of determining the optimal $M_k^*$ for each website is the non-cooperative game, and the utility maximizing problem of each website $k$ in terms of selling AD spaces can be formulated as follows:
\begin{align}
&\max_{M_k}V_k^A(M_k,\bm{M}_{-k},P_a)  = \max_{M_k}{\frac{M_k}{\sum_{j=1}^K {M_j}}}P_a  - C_kM_k\label{Problom_utility_webAD}\\
&\text{s.t.}~ M_k \ge 0.\label{Problom_utility_webAD Contraint}
\end{align}
for $k\in \mathcal{K}$.
 The first term in (\ref{Problom_utility_webAD}), i.e., ${\frac{M_k}{\sum_{j=1}^K {M_j}}}P_a $, represents the reward of video website $k$, and the second term in (\ref{Problom_utility_webAD}), i.e., $C_kM_k$, means the corresponding total ADs maintenance cost.
 The constraint in (\ref{Problom_utility_webAD Contraint}) indicates that the selling number of AD spaces for each video website $k$ is non-negative.

\textbf{Assumption}: \textit{we consider the case $\sum_{j\in\mathcal{K}\backslash\{k\}}M_j=0$, i.e., all other websites except website $k$ don't participate in the advertising budget game, then, the utility in (\ref{AD_Utility}) is reduced to $V_k^A(M_k,\bm{M}_{-k},P_a) = P_a  - C_kM_k$.
 In this case, there exists no optimal solution (equilibrium) for problem in (\ref{Problom_utility_webAD}) since the utility $V_k^A(M_k,\bm{M}_{-k},P_a)$ can be infinitely tending to $P_a$ when we choose a sufficiently small positive $ M_k$.
 Therefore, we here \textit{assume} that $\sum_{j\in\mathcal{K}\backslash\{k\}}M_j>0$ in this paper.}

 We note that the objective function in problem (\ref{Problom_utility_webAD}) is concave since ${\frac{\partial^2 V_k^A(M_k,\bm{M}_{-k},P_a)}{\partial M_k^2}}=-{\frac{2\sum_{j\in\mathcal{K}\backslash\{k\}}M_j}{\left({\sum_{j\in \mathcal{K}}M_j}\right)^3}}P_a <0$ holds, and the constraint in (\ref{Problom_utility_webAD Contraint}) is linear.
 Therefore, the problem in (\ref{Problom_utility_webAD}) is convex, and the Karush-Kuhn-Tucker (KKT) conditions
are the necessary and sufficient conditions of optimality \cite{Boyd2004}.
We thus derive the KKT conditions of problem in (\ref{Problom_utility_webAD}) as follows:
we let $\beta_k $, $k\in \mathcal{K}$, denote the the lagrange multiplier associated with the constraint in
(\ref{Problom_utility_webAD Contraint}).
The Lagrangian function of problem (\ref{Problom_utility_webAD}) with nonnegative $\beta_k $, $k\in \mathcal{K}$, can be given by:
 \begin{align}\label{L_Function}
\mathcal{L}(M_k, \boldsymbol{M}_{-k},P_a, \beta_k )& = {\frac{M_k}{\sum_{j=1}^K {M_j}}}P_a  - C_kM_k+ \beta_kM_k.
 \end{align}
 Denote $\beta_k^* $ as the optimal the optimal Lagrange multiplier for the corresponding dual problem.
 Therefore, the Karush-Kuhn-Tucker (KKT) conditions \cite{Boyd2004} are given by:
 \begin{align}\label{KKT1}
\frac{\partial\mathcal{L}}{\partial M_k^*}& = {\frac{\sum_{j\in\mathcal{K}\backslash\{k\}}M_j^*}{\left({\sum_{j\in \mathcal{K}}M_j^*}\right)^2}}P_a-C_k +\beta_k^* = 0,  ~~k \in \mathcal{K}.
 \end{align}
 Moreover, the non-negative Lagrange multiplier, i.e., $\beta_k^* \ge 0$,  should satisfy the the complementary slackness conditions \cite{Boyd2004}:
 \begin{align}
\beta_k^*M_k^* = 0, ~~k \in \mathcal{K}.\label{KKT2}.
 \end{align}
 Combining (\ref{KKT1})-(\ref{KKT2}) and simplifying, we have
  \begin{align}\label{M_k}
 M_k^*(\boldsymbol{M}^*_{-k},P_a) = \left[\sqrt{{\frac{P_a\sum_{j\in\mathcal{K}\backslash\{k\}}M_j^*}{C_k}}}-\sum_{j\in\mathcal{K}\backslash\{k\}}M_j^*\right]^+
 \end{align}
for $k\in \mathcal{K}$, where $[x]^+ \triangleq \max\{0, x\}$.

We can see from (\ref{M_k}) that: if $\sum_{j\in\mathcal{K}\backslash\{k\}}M_j^*< {\frac{P_a }{C_k}}$, we have $ M_k^*(\boldsymbol{M}^*_{-k},P_a)>0$; else if $\sum_{j\in\mathcal{K}\backslash\{k\}}M_j^*\ge {\frac{P_a }{C_k}}$, we have $ M_k^*(\boldsymbol{M}^*_{-k},P_a)=0$.
 Thus, we can further divide the optimal selling number of AD spaces $\boldsymbol{M}^*$ into two subsets.
 Specifically, one is the subset where the optimal strategy of each video website is participating in the advertising budget game, in other words, the optimal selling number of AD spaces is positive.
 We further denote this subset as $\bar{\mathcal{K}} \triangleq \{k \in \mathcal{K}|M_k^*>0\}$, and denote the corresponding size as $|\bar{\mathcal{K}}|$.
 The other is the subset where the optimal strategy of each video website is not participating in the advertising budget game, namely, the optimal selling number of the AD spaces is zero.
 Similarly, we denote this subset as $\mathcal{K} \backslash \bar{\mathcal{K}} \triangleq \{k\in \mathcal{K}|M_k^*=0\}$, and the corresponding size as $|\mathcal{K} \backslash \bar{\mathcal{K}}|$.

 In addition, we should notice that $M_k^*$ stated in (\ref{M_k}) is still a function of $\boldsymbol{M}^*_{-k}$ under $P_a$, and this coupled relationship does not provide the insight about the specific structure of the optimal solution.
Interestingly, we are fortunate to find out that: given subset $\bar{\mathcal{K}}$, we can derive the closed-form solution of $M_k^*$ under $P_a$, which is rigorously stated in the following Theorem.
\begin{Theo}\label{Optimal AD Space}
\rm{
Given subset $\bar{\mathcal{K}}$, the optimal selling number of AD spaces of video website $k$ under advertising budget $P_a$, i.e., $M_k^*(P_a)$, for problem in (\ref{Problom_utility_webAD}) is given by:
\begin{align}
M_k^*(P_a) =
\left\{ {\begin{array}{ll}\label{Opt_M_k_1}
{\frac{P_a(|\bar{\mathcal{K}}|-1)}{\sum_{j \in \bar{\mathcal{K}}} C_j}}\left( 1- {\frac{(|\bar{\mathcal{K}}|-1)C_k}{\sum_{j \in \bar{\mathcal{K}}} C_j}}\right), & \text{if} ~k\in \bar{\mathcal{K}}.\\
0,& \text{if}~ k \notin \bar{\mathcal{K}}.
\end{array}} \right.
 \end{align}
}
\end{Theo}
\begin{IEEEproof}
Please refer to Appendix \ref{Proof_Optimal AD Space}.
\end{IEEEproof}
Note that the advertising budget $P_a$ is given in stage I, thus, it can be considered as a constant during the determination of $M_k^*(P_a)$.
From (\ref{Opt_M_k_1}), we know that once $\bar{\mathcal{K}}$ is determined, the optimal solution for problem in (\ref{Problom_utility_webAD}), i.e., $M_k^*(P_a)$, is also obtained.
Therefore, our remaining work is to determine the subset $\bar{\mathcal{K}}$.
Before we proceed, we first provide some properties about $\bar{\mathcal{K}}$ based on the above arguments.
\begin{Prop}\label{K_Range}
\rm{
In the optimal solution for problem in (\ref{Problom_utility_webAD}), the size of the subset $\bar{\mathcal{K}}$ is no smaller than 2, i.e., $|\bar{\mathcal{K}}|\ge 2$.
}
\end{Prop}
\begin{IEEEproof}
Please refer to Appendix \ref{Proof_K_Range}.
\end{IEEEproof}

Combining Theorem \ref{Optimal AD Space} and Proposition \ref{K_Range}, we then have:
\begin{Prop}\label{C_Range}
\rm{
In the optimal solution for problem in (\ref{Problom_utility_webAD}), for each video website which belongs to subset $\bar{\mathcal{K}}$, its  maintenance cost of unit AD space $C_k$ satisfies $(|\bar{\mathcal{K}}|-1)C_k < \sum_{j\in\bar{\mathcal{K}}}C_j$.
}
\end{Prop}
\begin{IEEEproof}
Please refer to Appendix \ref{Proof_C_Range}.
\end{IEEEproof}
\begin{Prop}\label{K_Structure}
\rm{
In the optimal solution for problem in (\ref{Problom_utility_webAD}), if $\forall k \in \mathcal{K}$ such that $C_k \le \max_{j \in \bar{\mathcal{K}}} C_j$, then, $k \in \bar{\mathcal{K}}$.
}
\end{Prop}
\begin{IEEEproof}
Please refer to Appendix \ref{Proof_K_Structure}.
\end{IEEEproof}

Based on the insight provided by Proposition \ref{K_Range}-\ref{K_Structure}, we propose the following algorithm to compute the optimal subset $\bar{\mathcal{K}}$ for problem in (\ref{Problom_utility_webAD}).
\begin{algorithm}[htb]
\caption{The Optimal Subset $\bar{\mathcal{K}}$ Computation}
\begin{algorithmic}[1]\label{Opt_Subset_K}
\STATE Sort $C_k$, $k\in \mathcal{K}$ in ascending order: $C_{\vartheta_1}\le C_{\vartheta_2}\le \cdots \le C_{\vartheta_K}$;
\STATE Find $\vartheta_k = \argmax_{2 \le i\le K} \{(i-1)C_{\vartheta_i}<\sum_{j=1}^i C_{\vartheta_j}\}$;
\STATE Let $\bar{\mathcal{K}} \triangleq \{\vartheta_1,\vartheta_2,\cdots,\vartheta_k\}$.
\end{algorithmic}
\end{algorithm}

The first step in Algorithm \ref{Opt_Subset_K} is to sort the unit AD space maintenance cost $C_k$, $k \in \mathcal{K}$, in ascending order.
$\vartheta_i$, $i \in \mathcal{K}$ is the mapping index after the sorting operation for $C_k$, $k \in \mathcal{K}$.
For instance, we let $K = 3$, and further let $C_1 = 3$, $C_2 = 2$ and $C_3= 1$, respectively.
After the sorting operation, we have $C_3 < C_2 <C_1$, and thus $\vartheta_1 = 3$, $\vartheta_2 = 2$ and $\vartheta_3 = 1$.
In addition, Proposition \ref{K_Structure} indicates the structure of the optimal subset $\bar{\mathcal{K}}$, that is, each video website with the unit AD space cost no larger than the \textit{cost threshold} is the element of the optimal subset $\bar{\mathcal{K}}$, while the rest video websites belong to the subset $\mathcal{K} \backslash \bar{\mathcal{K}}$, where the \textit{cost threshold} is the maximal unit AD space cost satisfying the condition in proposition \ref{C_Range} for the ascending order AD space cost sequence, i.e., $\{C_{\vartheta_1},C_{\vartheta_2},\cdots,C_{\vartheta_K}\}$ in step 1.
Moreover, since $|\bar{\mathcal{K}}|\ge 2$ by Proposition \ref{K_Range}, it is natural that the index of video website with the maximal unit AD space cost (denoted by $\vartheta_k$) is an integer in $[\vartheta_2,\vartheta_K]$.
Therefore, through step 2, we can find the maximal cost index $\vartheta_k$, and thus the optimal subset is $\bar{\mathcal{K}} \triangleq \{\vartheta_1,\vartheta_2,\cdots,\vartheta_k\}$.

\subsection{Stage III: The optimal advertising Budget $P_a^*$}
In this subsection, we study the optimal advertising budget $P_a^*$ in stage I.
The advertiser determines its strategy through anticipating each video website's optimal AD space selling strategy $M_k^*$ and the optimal membership pricing strategy $p_k^*$, in stage II,  as well as the mobile user's optimal \textit{Non-Member} watching strategy probability ${\chi}_k^N(p_k^*)$ in stage III.
In addition, since only those video websites belong to subset $\bar{\mathcal{K}}$ participate in the game in stage I, it is natural to reformulate the utility function of the advertiser in (\ref{ADer_Utility}) as:
\begin{align}\label{ADer_Utility_New}
V^{AD}(P_a) = \sum_{k\in\bar{\mathcal{K}}} \omega_k G \phi_k N {\chi}_k^N(p_k^*) \tau_k(P_a, p_k^*,M_k^*) - P_a
\end{align}
where $p_k^*$ and $M_k^*$ are respectively computed in Theorem \ref{Optimal web price} and Theorem \ref{Optimal AD Space}, the optimal subset $\bar{\mathcal{K}}$ is determined in Algorithm \ref{Opt_Subset_K}.
\subsubsection{AD Spaces Saturation}
We can learn form Theorem \ref{Optimal AD Space} that $M_k^*$ increases linearly in $P_a$ for $k \in \bar{\mathcal{K}}$.
Therefore, the domain of $P_a$, i.e., $[0,+\infty)$, implies the domain of $M_k^*(P_a)$ is also $[0,+\infty)$.
Meanwhile, we should notice that the average total visiting number of website $k$'s advertising-containing videos per unit period is $\xi\phi_k N{\chi}_k^N(p_k^*)$, which indicates that: if $M_k^*(P_a) \in [0,\xi\phi_k N{\chi}_k^N(p_k^*))$, the specific advertisement's watching probability $\varphi_k(P_a, p_k^*,M_k^*)$ can be calculated by (\ref{ADWP}); if $M_k^*(P_a) \in [\xi\phi_k N{\chi}_k^N(p_k^*),+\infty)$, the specific advertisement's watching probability in (\ref{ADWP}) is replaced by $\varphi_k(P_a, p_k^*,M_k^*) = {\frac{1}{G}}$.
This is because the \textit{effective} total advertisement watching number is saturated when $M_k^*(P_a)\ge \xi\phi_k N{\chi}_k^N(p_k^*)$ due to the constraint of the average total visiting number, i.e., $\xi\phi_k N{\chi}_k^N(p_k^*)$.
Thus, when $M_k^*(P_a)\ge \xi\phi_k N{\chi}_k^N(p_k^*)$, $\varphi_k(P_a, p_k^*,M_k^*)$ reaches its maximum ${\frac{1}{G}}$ and increasing AD spaces $M_k^*$ cannot contribute for increasing $\varphi_k(P_a, p_k^*,M_k^*)$ anymore.
In the rest of this paper, if there exists $M_k^*(P_a)\ge \xi\phi_k N{\chi}_k^N(p_k^*)$, $k \in \bar{\mathcal{K}}$, we then call website $k$'s AD spaces are \textit{saturated}.
Based on the above arguments, we can naturally updated the \textit{AD Watching Probability} in (\ref{ADWatchPro}) as (\ref{ADWatchPro_New}).
\begin{figure*}[!htbp]
\begin{align}\label{ADWatchPro_New}
\bar{\tau}_k(P_a, p_k^*,M_k^*) =
\left\{ {\begin{array}{lr}
1 - \exp\left(-{\frac{M_k^*(P_a)}{G\phi_k N{\chi}_k^N(p_k^*)}}\right), &\text{if}~M_k^*(P_a)<\xi\phi_k N{\chi}_k^N(p_k^*); \\
1 - \exp(-{\frac{\xi}{G}}), &\text{if}~M_k^*(P_a) \ge \xi\phi_k N{\chi}_k^N(p_k^*).
\end{array}} \right. k\in \bar{\mathcal{K}}.
\end{align}
\end{figure*}

The advertiser's utility maximization problem can thus be formulated as follows:
\begin{align}\label{ADer_Utility_Prob}
&\max_{P_a}V^{AD}(P_a) = 
\max_{P_a}\sum_{k\in\bar{\mathcal{K}}} \omega_k G \phi_k N {\chi}_k^N(p_k^*)\bar{ \tau}_k(P_a, p_k^*,M_k^*) - P_a\\
&\text{s.t.}~P_a\ge 0.
\end{align}
\subsubsection{Advertising Budget Saturation Threshold}
Before we  proceed, we first provide a definition about the corresponding advertising budget threshold value $P_a$ when the website's AD spaces are \textit{saturated}.
\begin{Def}\label{P_a^T}
We define $P_a^{T,k} = P_a$ as the \textit{advertising budget saturation threshold} of video website $k$, $k\in \bar{\mathcal{K}}$, if $M^*_k(P_a) = \xi\phi_k N{\chi}_k^N(p_k^*)$, i.e., video website $k$'s AD spaces are saturated.
\end{Def}
Note that there are $|\bar{\mathcal{K}}|$ video websites in subset $\bar{\mathcal{K}}$, thus, there exists total $|\bar{\mathcal{K}}|$ \textit{advertising budget saturation thresholds}.
We further denote $P_a^{T,\max} \triangleq \max\{P_a^{T,1}, P_a^{T,2}, \cdots, P_a^{T,|\bar{\mathcal{K}}|}\}$ as the \textit{maximum advertising budget saturation threshold}, and denote $P_a^{T,\min} \triangleq \min\{P_a^{T,1}, P_a^{T,2}, \cdots, P_a^{T,|\bar{\mathcal{K}}|}\}$ as the \textit{minimum advertising budget saturation threshold}, respectively.
We sort these $|\bar{\mathcal{K}}|$ \textit{advertising budget saturation threshold} values in ascending order, and denote it by $\{P_a^{T,\alpha_1},P_a^{T,\alpha_2},\cdots, P_a^{T,\alpha_{|\bar{\mathcal{K}}|}}\}$, where $\alpha_k$, $k\in \bar{\mathcal{K}}$, is the mapping index after sorting operation.
Naturally, we have $P_a^{T,\min} = P_a^{T,\alpha_1} \le P_a^{T,\alpha_2} \le \cdots \le P_a^{T,\alpha_{|\bar{\mathcal{K}}|}} =P_a^{T,\max}$.
Moreover, if $P_a = P_a^{T,\alpha_k}$, then, the AD spaces of all the websites in set $\{\alpha_1, \cdots, \alpha_k\}$ are \textit{saturated}.
\subsubsection{The optimal advertising Budget $P_a^*$}
Based on the above argument, we can have the following lemma which describes the upper bound of $P_a^*$ for problem in (\ref{ADer_Utility_Prob}).
\begin{Lem}\label{P_a_UpperBound}
The optimal solution of problem in (\ref{ADer_Utility_Prob}) $P_a^*$ satisfies $P_a^* \le P_a^{T,\max}$.
\end{Lem}
\begin{IEEEproof}
Please refer to Appendix \ref{Proof P_a_UpperBound}
\end{IEEEproof}

Combining the \textit{advertising budget saturation threshold} and Lemma \ref{P_a_UpperBound}, we can equivalently reformulate the objective function in (\ref{ADer_Utility_Prob}) as in (\ref{ADer_Utility_New1}).
\begin{figure*}
\begin{align}\label{ADer_Utility_New1}
V^{AD}(P_a) =
\left\{ {\begin{array}{l}
\sum\limits_{k\in\bar{\mathcal{K}}} \omega_{\alpha_k} G \phi_{\alpha_k} N {\chi}_{\alpha_k}^N(p_{\alpha_k}^*)\tau_{\alpha_k}(P_a, p_{\alpha_k}^*,M_{\alpha_k}^*) - P_a,\\
0 \le P_a \le P_a^{T,\min}; \\
\sum\limits_{k=2}^{{|\bar{\mathcal{K}}|}} \omega_{\alpha_k} G \phi_{\alpha_k} N {\chi}_{\alpha_k}^N(p_{\alpha_k}^*)\tau_{\alpha_k}(P_a, p_{\alpha_k}^*,M_{\alpha_k}^*) +
\omega_{\alpha_1} G \phi_{\alpha_1} N {\chi}_{\alpha_1}^N(p_{\alpha_1}^*)(1 - \exp(-{\frac{\xi}{G}}))- P_a,\\
P_a^{T,\min} \le P_a \le P_a^{T,\alpha_2};\\
 \vdots  \\
 \omega_{\alpha_{|\bar{\mathcal{K}}|}} G \phi_{\alpha_{|\bar{\mathcal{K}}|}} N {\chi}_{\alpha_{|\bar{\mathcal{K}}|}}^N(p_{\alpha_{|\bar{\mathcal{K}}|}}^*)\tau_{\alpha_{|\bar{\mathcal{K}}|}}(P_a, p_{\alpha_{|\bar{\mathcal{K}}|}}^*,M_{\alpha_{|\bar{\mathcal{K}}|}}^*)+
\sum\limits_{k=1}^{|\bar{\mathcal{K}}|-1}\omega_{\alpha_k} G \phi_{\alpha_k} N {\chi}_{\alpha_k}^N(p_{\alpha_k}^*)(1 - \exp(-{\frac{\xi}{G}}))- P_a,\\
 P_a^{T,\alpha_{|\bar{\mathcal{K}}|-1}} \le P_a \le P_a^{T,\max}.
\end{array}} \right.
\end{align}
\end{figure*}
It is not difficult to find: 1) function $V^{AD}(P_a) $ in (\ref{ADer_Utility_New1}) is with piecewise structure; and 2) function $V^{AD}(P_a) $ in (\ref{ADer_Utility_New1}) is continuous since we can verified that each utility function in (\ref{ADer_Utility_New1}) is continuous in its own feasible domain, and $V^{AD}(P_a^{T,\alpha_k-})=V^{AD}(P_a^{T,\alpha_k})=V^{AD}(P_a^{T,\alpha_k+})$, $k \in \bar{\mathcal{K}}$, by (\ref{ADWatchPro_New}) and (\ref{ADer_Utility_New1}), where $x^- = \lim_{\epsilon \to 0^+} x-\epsilon$ and $x^+ = \lim_{\epsilon \to 0^+} x+\epsilon$.
Therefore, to determine the optimal solution of problem in (\ref{ADer_Utility_Prob}) is equivalent to find out a specific $P_a$ in one of ${|\bar{\mathcal{K}}|}$ feasible domains (i.e., $[0,P_a^{T,\min}], \cdots, [P_a^{T,\alpha_{|\bar{\mathcal{K}}|-1}},P_a^{T,\max}]$) to maximize its corresponding objective function in (\ref{ADer_Utility_New1}), meanwhile, the corresponding objective function value is larger than the maximum value of any other objective function in (\ref{ADer_Utility_New1}) in its own feasible domain.

The remaining work is to determine the specific expression form of the objective function in (\ref{ADer_Utility_New1}), and to calculate its optimal solution of its corresponding objective function maximization problem within the feasible domain.
Before we proceed, we first define $\mathcal{S}$ as the set of those websites such that their AD spaces are \textit{saturated}.
In addition, the ${|\bar{\mathcal{K}}|}$ \textit{advertising budget saturation thresholds}' relationship, i.e., $P_a^{T,\min} = P_a^{T,\alpha_1} \le P_a^{T,\alpha_2} \le \cdots \le P_a^{T,\alpha_{|\bar{\mathcal{K}}|}} =P_a^{T,\max}$, determines that the set $\mathcal{S}$ can only be one element of $\bar{\mathcal{S}} \triangleq \{\emptyset,\{1\}, \{1,2\},\cdots, \{1,2,\cdots,|\bar{\mathcal{K}}-1|\} \}$, where $\bar{\mathcal{S}}$ denotes the feasible domain of set $\mathcal{S}$.
For notation simplification, we respectively define $\bar{\mathcal{S}}_{0} \triangleq {\emptyset}$ and $\bar{\mathcal{S}}_k \triangleq \{1,\cdots,k\}$, where $k \in \{1,\cdots,|\bar{\mathcal{K}}-1|\}$.
Therefore, $\bar{\mathcal{S}} \triangleq \{\bar{\mathcal{S}}_0, \bar{\mathcal{S}}_1, \cdots, \bar{\mathcal{S}}_{|\bar{\mathcal{K}}|-1}\}$.
Then, for any $\mathcal{S} \in \bar{\mathcal{S}}$, we consider an optimization problem:
\begin{align}\label{New_Opt}
\max_{P_a} F(P_a)&=
\max_{P_a} \sum_{k\in\mathcal{S}} \omega_{\alpha_k} G \phi_{\alpha_k} N {\chi}_{\alpha_k}^N(p_{\alpha_k}^*)\left(1 - \exp\left(-{\frac{\xi}{G}}\right)\right)\nonumber\\
&+ \sum_{k \in \bar{\mathcal{K}}\backslash \mathcal{S}} \omega_{\alpha_k} G \phi_{\alpha_k} N {\chi}_{\alpha_k}^N(p_{\alpha_k}^*)\tau_{\alpha_k}(P_a, p_{\alpha_k}^*,M_{\alpha_k}^*) -P_a.\\
&\text{s.t.}~P_a\ge 0.
\end{align}
Problem in (\ref{New_Opt}) can be considered as the problem in (\ref{ADer_Utility_Prob}) under given set $\mathcal{S}$ with the feasible domain relaxing to $[0, +\infty)$.
Specifically, no matter $\mathcal{S}$ equals to any element in $\bar{\mathcal{S}}$, the objective function in (\ref{New_Opt}) is the same as the corresponding objective function in (\ref{ADer_Utility_New1}), and the only difference between these two problems are the feasible domains.
For instance, we assume $\mathcal{S} =\bar{\mathcal{S}}_1 =\{1\}$, the feasible domain of problem in (\ref{New_Opt}) is $[0, +\infty)$, while the feasible domain of problem in (\ref{ADer_Utility_Prob}) is $(P_a^{T,\alpha_{1}},P_a^{T,\alpha_{2}}]$.

Next, we will give a lemma which characterizes the relationship among different optimal solutions for problem in (\ref{New_Opt}) under different set $\mathcal{S}$.
\begin{Lem}\label{Opt_P_a_relationship}
For $\mathcal{S}, \mathcal{S}' \in \bar{\mathcal{S}}$, if $\mathcal{S} \subseteq \mathcal{S}'$, then, the optimal solution for problem in (\ref{New_Opt}) satisfies $P_{a,\mathcal{S}}^*\ge P_{a,\mathcal{S}'}^*$, where $P_{a,\mathcal{S}}^*$ and $P_{a,\mathcal{S}'}^*$ are the optimal solution corresponds to $\mathcal{S}$ and $\mathcal{S}'$, respectively.
\end{Lem}
\begin{IEEEproof}
Please refer to Appendix \ref{Proof_Opt_P_a_relationship}.
\end{IEEEproof}

Based on the above arguments, we will propose the algorithm to calculate the optimal solution $P_a^*$ for problem in (\ref{ADer_Utility_Prob}), the rigorous description is stated as follows:
\begin{algorithm}[htb]
\caption{The Optimal Advertising Budget $P_a^*$ Computation}
\begin{algorithmic}[1]\label{Opt_P_a_algorithm}
\STATE Let $\mathcal{S} = \bar{\mathcal{S}} _0=\emptyset$ and compute the corresponding optimal solution $P_{a,\mathcal{S}= \bar{\mathcal{S}} _0}^*$ for problem in (\ref{New_Opt}).
\STATE If $P_{a,\mathcal{S}= \bar{\mathcal{S}} _0}^* \le P_a^{T,\min}$, then, $P_a^* = P_{a,\mathcal{S}= \bar{\mathcal{S}} _0}^*$; else, goto step 3.
\STATE Let $\mathcal{S} = \bar{\mathcal{S}}_k$ for all $k \in \{1,\cdots,|\bar{\mathcal{K}}-1|\}$, and compute the corresponding optimal solution $P_{a,\mathcal{S}= \bar{\mathcal{S}}_k}^*$ for problem in (\ref{New_Opt}).
\STATE Find out $k^*=\argmax_{k\in \{1,\cdots,|\bar{\mathcal{K}}-1|\}}   {P_{a,\mathcal{S}= \bar{\mathcal{S}}_k}^*\ge P_a^{T,\alpha_k}}$. If $k^*$ exists, $P_a^* = \min\{P_{a,\mathcal{S}= \bar{\mathcal{S}}_{k^*}}^*, P_a^{T,\alpha_{{k^*}+1}}\}$; else, $P_a^* = P_a^{T,\min}$.
\end{algorithmic}
\end{algorithm}

The key of this algorithm is to 1) solve the optimal solution $P_{a,\mathcal{S}}^*$, $\mathcal{S} \in \bar{\mathcal{S}}$, for problem in (\ref{New_Opt}), and 2) to determine the relationship between $P_{a,\mathcal{S}}^*$ and the corresponding objective function's feasible domain in (\ref{ADer_Utility_New1}).
In specific, we first let $\mathcal{S} = \bar{\mathcal{S}} _0=\emptyset$, and obtain $P_{a,\mathcal{S}}^*$ by (\ref{1oderde}). If $P_{a,\mathcal{S}= \bar{\mathcal{S}} _0}^* \le P_a^{T,\min}=P_a^{T,\alpha_1}$, then, Problem in (\ref{New_Opt}) and problem in (\ref{ADer_Utility_Prob}) is equivalent, thus $P_a^* = P_{a,\mathcal{S}= \bar{\mathcal{S}} _0}^*$; else, there exists at least one video website such that the AD spaces are \textit{saturated} and the optimal advertising budget $P_a^*$ satisfies $P_a^* \ge  P_a^{T,\min}$ due to the concavity and continuity of $V^{AD}(P_a)$ in (\ref{ADer_Utility_New1}). Specifically, if there does not exists an index $k^* \in \{1,\cdots,|\bar{\mathcal{K}}-1|\}$, i.e., $P_{a,\mathcal{S}= \bar{\mathcal{S}} _0}^* \ge P_a^{T,\min}> P_{a,\mathcal{S}= \bar{\mathcal{S}} _1}^*\ge \cdots \ge P_{a,\mathcal{S}= \bar{\mathcal{S}} _{|\bar{\mathcal{K}}-1|}}^*$, in this case, according to the concavity of $V^{AD}(P_a)$ in (\ref{ADer_Utility_New1}), $V^{AD}(0<P_a \le P_a^{T,\min})$ achieves its maximum value at $P_a^{T,\min}$ and $V^{AD}(P_a^{T,\alpha_k}  \le P_a < P_a^{T,\alpha_{k+1}} )$, $k \in \{1,\cdots,|\bar{\mathcal{K}}-1|\}$ achieves its maximum value at $P_a^{T,\alpha_k}$. Since $V^{AD}(P_a)$ in (\ref{ADer_Utility_New1}) is continuous, we have $P_a^* = P_a^{T,\min}$.
 If the index $k^*$ exists, according to lemma \ref{Opt_P_a_relationship}, it can guarantee that ${P_{a,\mathcal{S}= \bar{\mathcal{S}}_k}^*\ge P_a^{T,\alpha_k}}$ for $\alpha_k \le \alpha_{k^*}$ and ${P_{a,\mathcal{S}= \bar{\mathcal{S}}_k}^*< P_a^{T,\alpha_k}}$ for $\alpha_k > \alpha_{k^*}$. similar to the above analysis, we have $P_a^* = \min\{P_{a,\mathcal{S}= \bar{\mathcal{S}}_{k^*}}^*, P_a^{T,\alpha_{{k^*}+1}}\}$.
\begin{Theo}\label{Algo2Opt}
Algorithm \ref{Opt_P_a_algorithm} can find the optimal Advertising Budget $P_a^*$ for Problem in (\ref{ADer_Utility_Prob}).
\end{Theo}
\begin{IEEEproof}
Please refer to Appendix \ref{Proof_Algo2Opt}.
\end{IEEEproof}

\section{Numerical Results}
In this section, we provide the numerical results to verify the impacts of the varies of different parameters' values on $V^{AD}$, $V_k^{VW}$, $P_a^*$, $p_k^*$, $M_k^*$ and the key indicators.
\subsection{The impact of different mean values of $\sigma_k$}\label{sigma}
In this subsection, we show the impact of the \textit{VIP-Member} price coefficient $\sigma_k$ on the membership probability (i.e., ${\chi}_k^V$, ${\chi}_k^R$ and ${\chi}_k^N$), the optimal membership price (i.e., $p_k^*$ and $\sigma_kp_k^*$), the optimal advertising budget (i.e., $P_a^*$) and all utilities (i.e., $V_k^{M^*}$, $V_k^{A^*}$, $V_k^{{VW}^*}$ and $V^{{AD}^*}$).
We choose $K = 10$, $\theta_{\max} = 1$, $Q_k^V = 40$, $Q_k^R = 20$, $Q_k^N = 10$, $C_k = k, k = \{1,2,\cdots,K\}$, $\xi = 10$, $\gamma = 0.5$, $N = 5000$, $G = 5$ and $\omega_k = 100$.
In addition, we let the mean of $\sigma_k$ vary from 3 to 150, i.e., $\sigma_k \in [\sigma_k^T, 50\sigma_k^T]$.

\begin{figure}[htbp]
\centering
\includegraphics[width=0.5\textwidth]{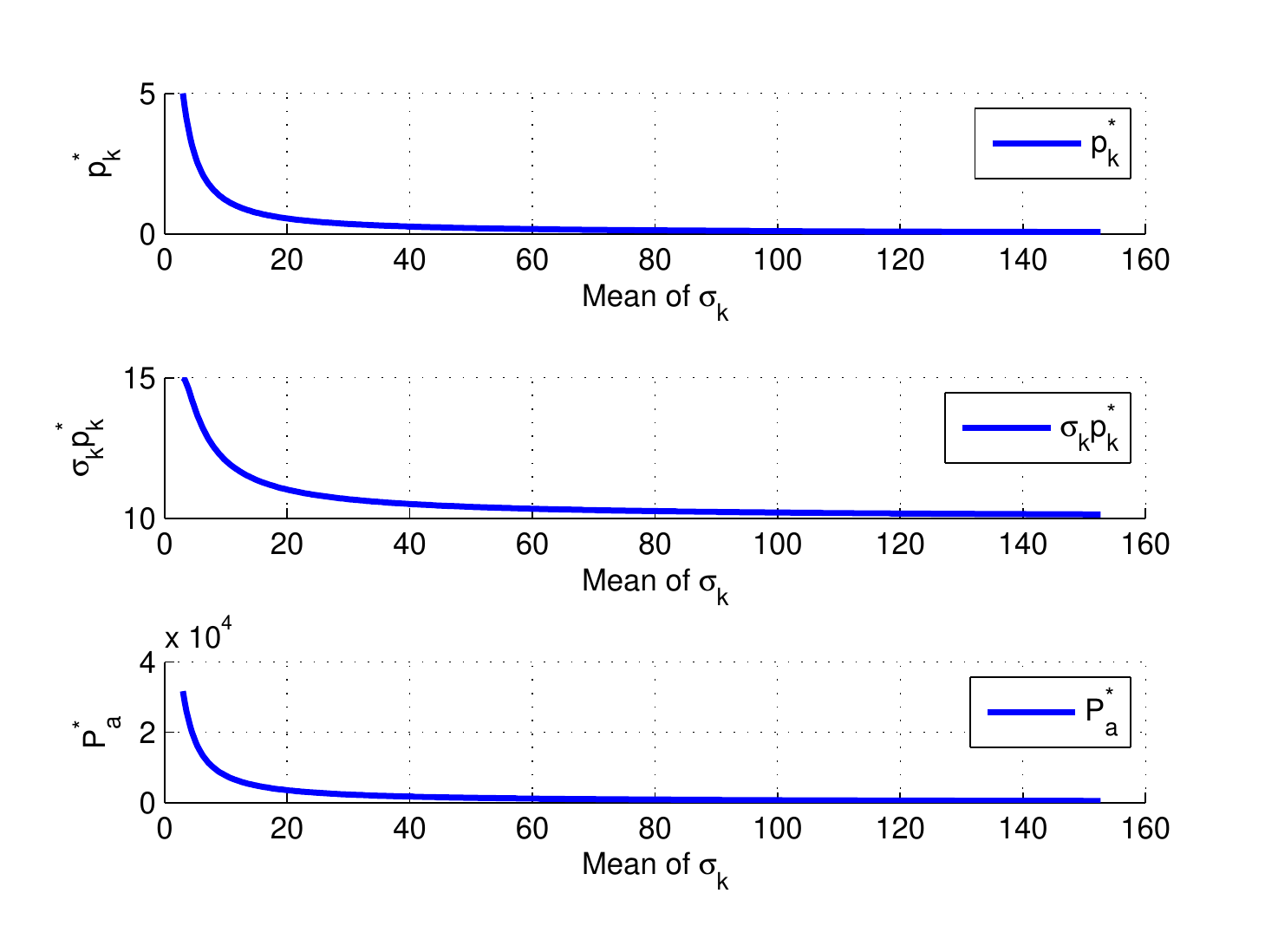}
\caption{The impacts of different mean values of $\sigma_k$ on the membership price and the advertising budget.}
\label{SigmaVSprice}
\end{figure}

Fig. \ref{SigmaVSprice} presents the impacts of different mean values of $\sigma_k$ on the optimal \textit{Regular-Member} price $p_k^*$, the optimal \textit{VIP-Member} price $\sigma_k p_k^*$ and the optimal advertising budget $P_a^*$, respectively.
We can see that both \textit{Regular-Member} price and \textit{VIP-Member} price are decreasing in $\sigma_k$.
In addition, when $\sigma_k$ is sufficiently large, we have $\lim\limits_{\sigma_k \to \infty}p_{k}^* =\lim\limits_{\sigma_k \to \infty} {\frac{\theta_{\max}( Q_k^V- Q_k^R)}{2(\sigma_k-1)}} = 0$ and $\lim\limits_{\sigma_k \to \infty}\sigma_k p_{k}^* = {\frac{\theta_{\max}( Q_k^V- Q_k^R)}{2}} = 10$, which also explain that \textit{Regular-Member} price and \textit{VIP-Member} price respectively converge to 0 and 10 when $\sigma_k$ approaches to 150. In addition, we also find $P_a^*$ is decreasing in the mean values of $\sigma_k$.

\begin{figure}[htbp]
\centering
\includegraphics[width=0.5\textwidth]{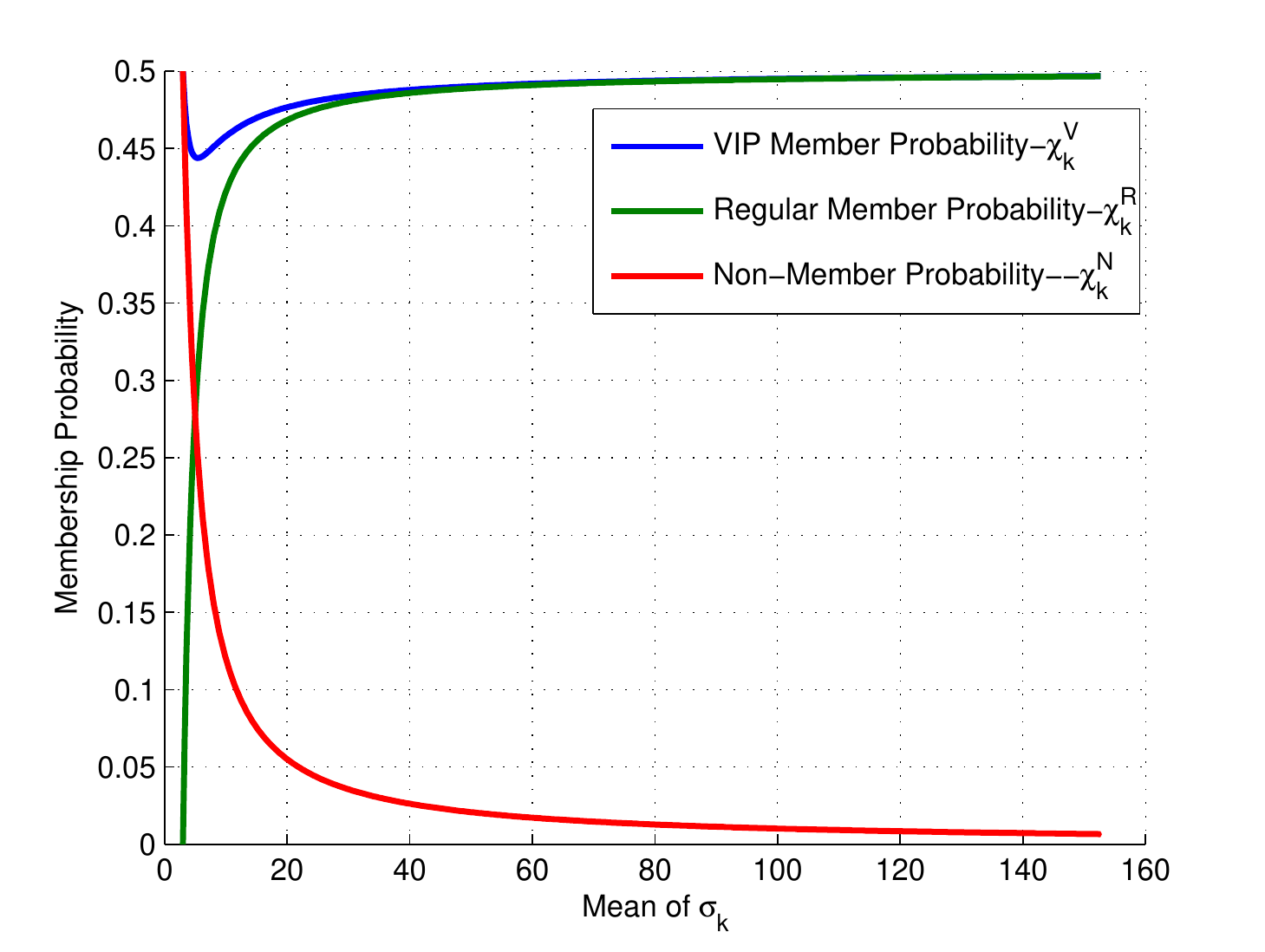}
\caption{The impacts of different mean values of $\sigma_k$ on the membership probability.}
\label{SigmaVSProbability}
\end{figure}

Fig. \ref{SigmaVSProbability} shows the membership probability of mobile user under different mean values of $\sigma_k$.
It can be easily observed that ${\chi}_k^V$  first decreases and then increases in $\sigma_k$, ${\chi}_k^R$ increases in $\sigma_k$ and ${\chi}_k^N$ decreases in $\sigma_k$.
Moreover, when $\sigma_k = \sigma_k^T = 3$, $p_k^*$ in  (\ref{Opt_MemPrice}) can be simplified as $p_k^* = \theta_{\max}(Q^R_k-Q^N_k)/2$, and thus we have $\theta_k^{T_{12}} = \theta_k^{T_{23}}=\theta_k^{T_{13}}= \theta_{\max}/2=1/2$ by substituting $p_k^* = \theta_{\max}(Q^R_k-Q^N_k)/2$ into definition \ref{Threshold1}-\ref{Threshold3}.
According to (\ref{Prob_VIP})-(\ref{Prob_Non}), we have ${\chi}_k^V={\chi}_k^N=1/2$ and ${\chi}_k^R=0$, that is to say, each mobile user's strategy is to be \textit{Non-Member} or \textit{VIP-Member} of website $k$, which is in accordance with the case in (\ref{Watching_Strategy_New}).
When $\sigma_k$ is sufficiently large, i.e., $\sigma_k\rightarrow \infty$, we have $\lim\limits_{\sigma_k \to \infty}p_{k}^* =\lim\limits_{\sigma_k \to \infty} {\frac{\theta_{\max}( Q_k^V- Q_k^R)}{2(\sigma_k-1)}} = 0$, $\lim\limits_{\sigma_k \to \infty}\theta_k^{T_{12}} ={\frac{1}{2}}$, $\lim\limits_{\sigma_k \to \infty}\theta_k^{T_{23}} =\lim\limits_{\sigma_k \to \infty}{\frac{(Q^V_k-Q^R_k)}{ 2(\sigma_k-1)(Q^R_k-Q^N_k)}}=0$, and further $\lim\limits_{\sigma_k \to \infty}{\chi}_k^R = \lim\limits_{\sigma_k \to \infty}{\chi}_k^V = {\frac{1}{2}}$, $\lim\limits_{\sigma_k \to \infty}{\chi}_k^N = 0$. Therefore, when $\sigma_k$ approaches to 150, ${\chi}_k^V$ and ${\chi}_k^R$ converge to ${\frac{1}{2}}$, and ${\chi}_k^N$ converges to 0, respectively. In other words, each mobile user's strategy is to be \textit{VIP-Member} or \textit{Regular-Member} of website $k$.

\begin{figure}[htbp]
\centering
\includegraphics[width=0.5\textwidth]{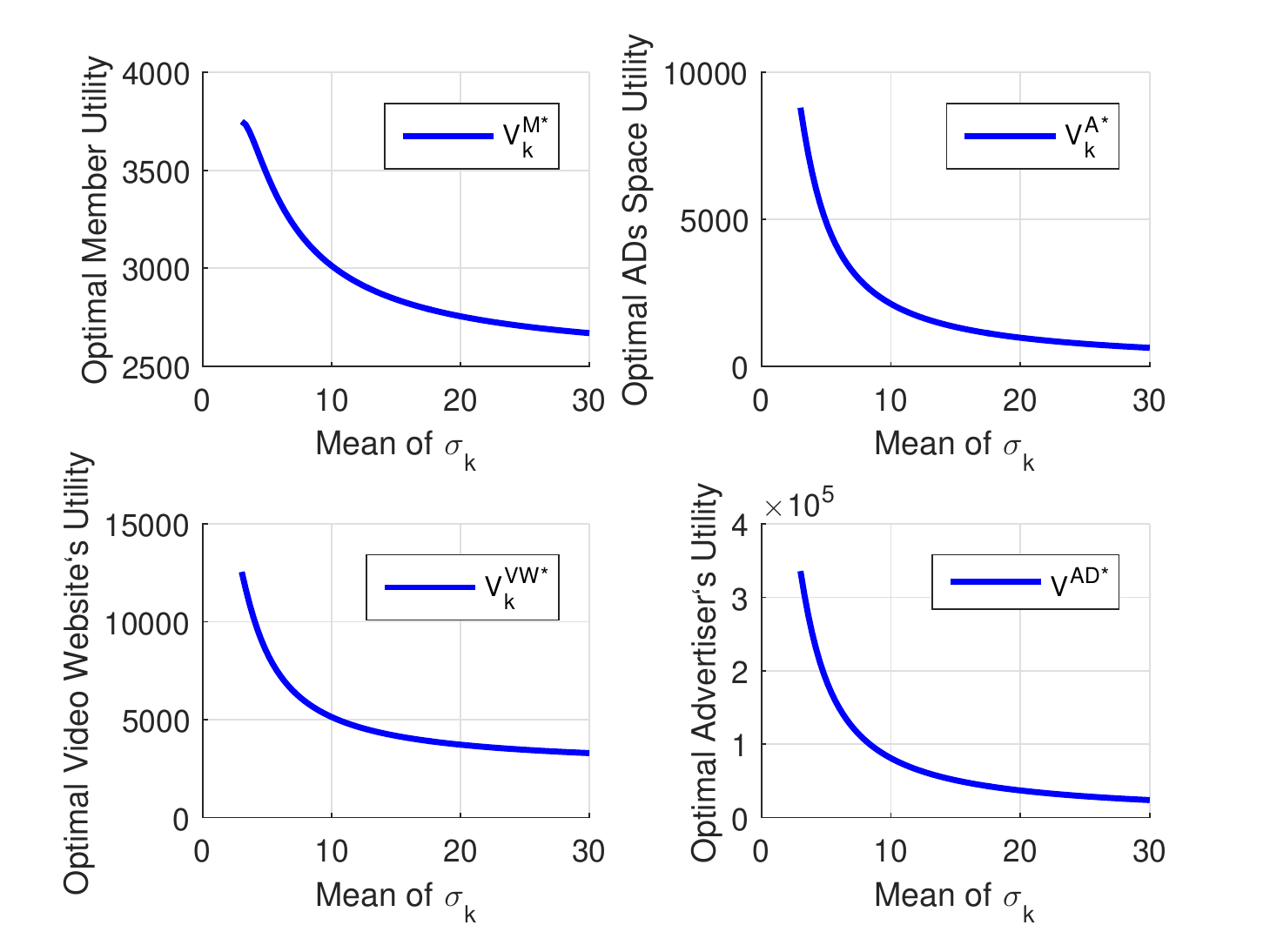}
\caption{The impacts of different mean values of $\sigma_k$ on each utility.}
\label{SigmaVSUtility}
\end{figure}

Fig. \ref{SigmaVSUtility} illustrates the impacts of different mean values of $\sigma_k$ on the membership utility ($V_k^{M^*}$), the AD space utility ($V_k^{A^*}$), the video website's utility ($V_k^{{VW}^*}$) and the advertiser's utility ($V^{{AD}^*}$), respectively. Here, we let $\sigma_k \in [3, 30]$. Firstly, we can observe that $V_k^M$ is decreasing in $\sigma_k$. This is because although we can observe from Fig.\ref{SigmaVSProbability} that ${\chi}_k^V$ and ${\chi}_k^R$ are increasing in $\sigma_k$, both $p_k^*$ and $\sigma_kp_k^*$ are decreasing in $\sigma_k$ from Fig. \ref{SigmaVSprice} and the influences of them on $V_k^{M^*}$ are dominant, thus, $V_k^{M^*}$ is decreasing in $\sigma_k$ by (\ref{Utility_Member}). Secondly, substituting (\ref{Opt_M_k_1}) and (\ref{Sum_all_M_k}) into (\ref{AD_Utility}) and simplifying, we have $V_k^{A^*} = P_a^*\left( 1- {\frac{(|\bar{\mathcal{K}}|-1)C_k}{\sum_{j \in \bar{\mathcal{K}}} C_j}}\right)^2$, which implies that $V_k^{A^*}$ is decreasing in $\sigma_k$ since $P_a^*$ decreases in $\sigma_k$ from Fig.\ref{SigmaVSprice}. Thirdly, it is nature that $V_k^{{VW}^*}$ decreases in $\sigma_k$ since $V_k^{VW} =V_k^{M} +V_k^{A}$. Fourthly, we can find that $V^{{AD}^*}$ is decreasing in $\sigma_k$. This is because ${\chi}_k^N$ decreases in $\sigma_k$ from Fig.\ref{SigmaVSprice}, and the impact of $\sum_{k\in\bar{\mathcal{K}}} \omega_k G \phi_k N {\chi}_k^N(p_k^*)\bar{ \tau}_k(P_a, p_k^*,M_k^*)$ in (\ref{ADer_Utility_Prob}) is dominant, thus, $V^{AD}$ decreases in $\sigma_k$.

In addition, since the utilities of the video website and the advertiser, i.e., $V_k^{{VW}^*}$ and $V^{{AD}^*}$, are decreasing in $\sigma_k$ from Fig.\ref{SigmaVSUtility}, we can conclude that the increase of $\sigma_k$ is not beneficial to improve the utilities of the video website and the advertiser.

\subsection{The impact of different $\gamma$}
In this section, we demonstrate the impact of video website's popularity concentrate level parameter $\gamma$ on the optimal adverting budget (i.e., $P_a^*$), the optimal AD spaces selling number (i.e., $M_k^*$), and each utility (i.e., $V_k^M$, $V_k^A$, $V_k^{VW}$ and $V^{AD}$), respectively.
Here, we let $\sigma_k = 1.25\sigma_k^T$, $\gamma \in [0,1]$.
 We respectively consider two AD cost parameter settings: $C_k = k$ and $C_k = K -k +1$, where $k = \{1,2,\cdots,K\}$.
The remaining parameter settings are the same as section \ref{sigma}.

\begin{figure}[htbp]
\centering
\includegraphics[width=0.5\textwidth]{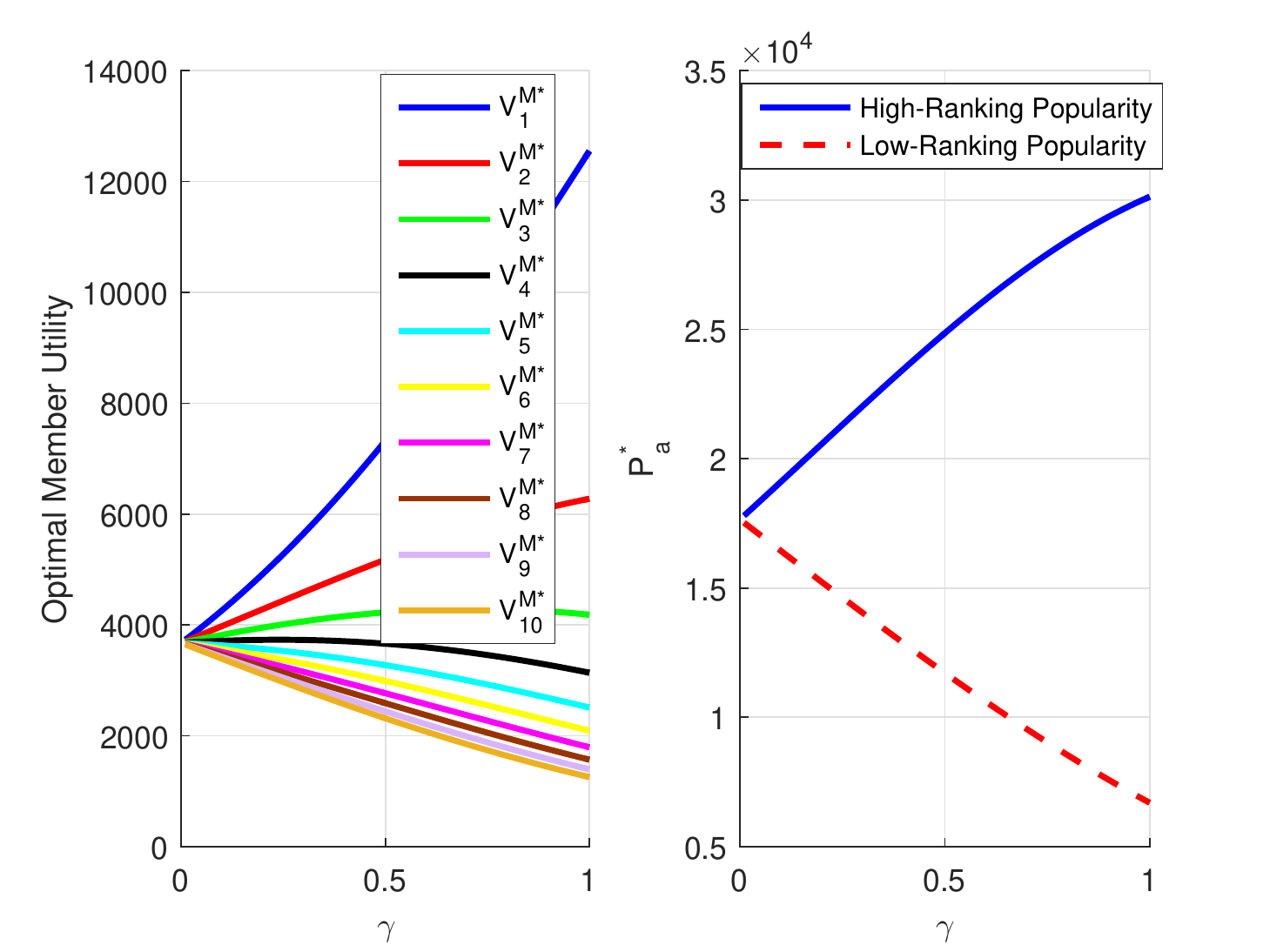}
\caption{The impacts of different values of $\gamma$ on $V_k^{{M}^*}$ and $P_a^*$.}
\label{GammaVSM-ADutility}
\end{figure}

Fig.\ref{GammaVSM-ADutility} depicts the impact of $\gamma$ on $V_k^{M^*}$ and $P_a^*$. Since parameter $\gamma$ describes the \textit{concentration level} of the video website's popularity distribution, with increasing of $\gamma$, the popularity concentrates on the high-ranking video websites (i.e., the websites with small index number). In other words,  the popularity of high-ranking video website increases in $\gamma$, and the popularity of low-ranking video website decreases in $\gamma$. According to (\ref{Utility_Member}), we know that the monotonicity of $V_k^M$ is the same with that of $\phi_k$. Meanwhile, we can see from the left sub-fig in Fig. \ref{GammaVSM-ADutility} that only $\phi_1$ (or $V_1^{M^*}$) and $\phi_2$ (or $V_2^{M^*}$) increase in $\gamma$; $\phi_3$ (or $V_3^{M^*}$) first increases and then decreases in $\gamma$; $\phi_k$ (or $V_k^{M^*}$), $k = 3,\cdots,K$ decreases in $\gamma$.

 In addition, we note that the subset $\bar{\mathcal{K}}$ only depends on the AD cost $C_k$ in $\mathcal{K}$. According to those two AD cost settings in the experiment, we have $\bar{\mathcal{K}}$ equal to $\{1, 2\}$ for case $C_k = k$ and $\{9, 10\}$ for case $C_k = K -k +1$, respectively. Meanwhile, based on the results of the left sub-fig in Fig. \ref{GammaVSM-ADutility}, we know that $\phi_k$, $k \in \{1, 2\}$ increases in $\gamma$ and we can say that the websites participating in the game are with high-ranking popularity.  For $\bar{\mathcal{K}} = \{9, 10\}$, $\phi_k$, $k \in \bar{\mathcal{K}}$ decreases in $\gamma$ and we can say that the websites participating in the game are with low-ranking popularity. Therefore, for $k \in \{1, 2\}$, $B_k = G\phi_k N{\chi}_k^N(p_k^*)$ is increasing in $\gamma$, and further $P_a^*$ is increasing in $\gamma$ by (\ref{first_order}), which is shown in the right sub-fig of Fig. \ref{GammaVSM-ADutility}. For $k \in \{9, 10\}$, through the similar analysis to the case $k \in \{1, 2\}$, $P_a^*$ is decreasing in $\gamma$, as shown in the dashed line of Fig. \ref{GammaVSM-ADutility}.

\begin{figure}[htbp]
\centering
\includegraphics[width=0.5\textwidth]{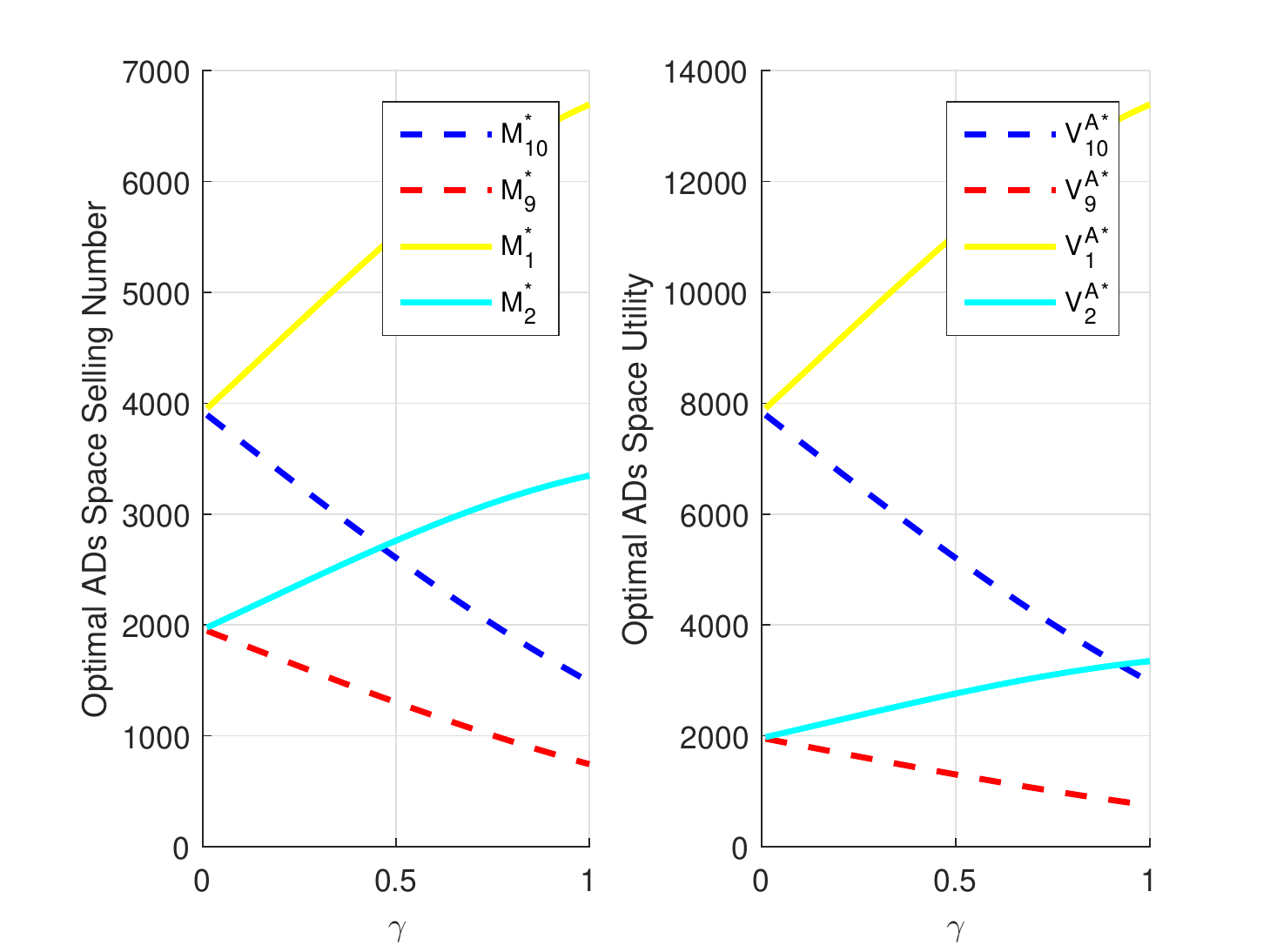}
\caption{The impacts of different values of $\gamma$ on $M_k^*$ and $V_k^{A^*}$.}
\label{GammaVSMember-ADutility}
\end{figure}

 Fig.\ref{GammaVSMember-ADutility} shows the impact of $\gamma$ on $M_k^*$ and  $V_k^{A^*}$.  For $k \in \{1, 2\}$, since $P_a^*$ is increasing in $\gamma$ from Fig.\ref{GammaVSM-ADutility}, we can know that $M_1^*$ and $M_2^*$ are also increasing in $\gamma$ by (\ref{Opt_M_k_1}), as shown in the left sub-fig in Fig.\ref{GammaVSMember-ADutility}. Meanwhile, similar to the analysis of Fig.\ref{GammaVSM-ADutility}, since $V_k^{A} = P_a\left( 1- {\frac{(|\bar{\mathcal{K}}|-1)C_k}{\sum_{j \in \bar{\mathcal{K}}} C_j}}\right)^2$, $V_1^{A^*}$ and $V_2^{A^*}$ increase in $\gamma$, as shown in the right sub-fig in Fig.\ref{GammaVSMember-ADutility}. For $k \in \{9, 10\}$, we can see that $M_9^*$, $M_{10}^*$, $V_9^{A^*}$ and $V_{10}^{A^*}$ are decreasing in $\gamma$, as shown in the dashed lines.

\begin{figure}[htbp]
\centering
\includegraphics[width=0.5\textwidth]{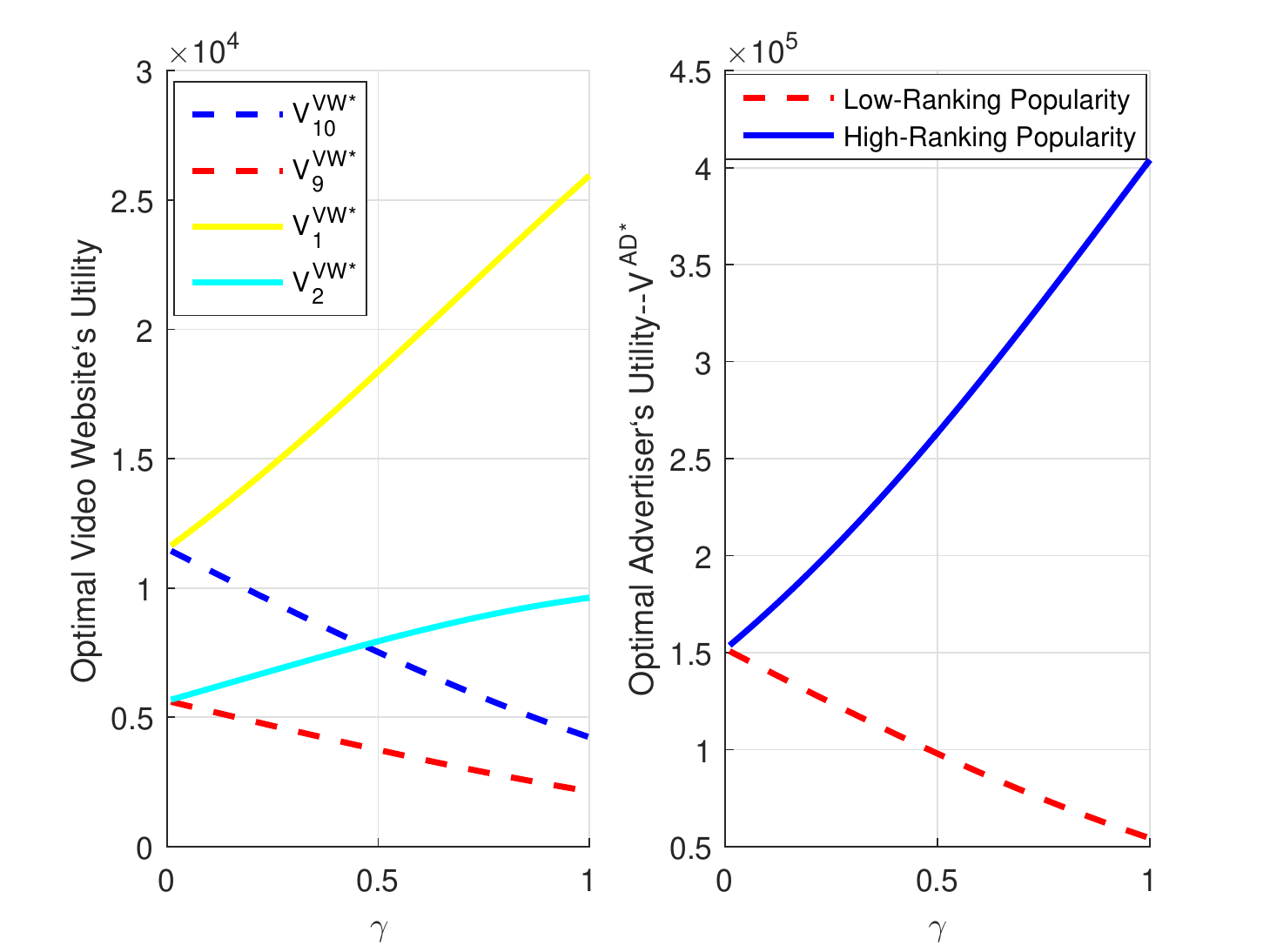}
\caption{The impacts of different values of $\gamma$ on $V_k^{{VW}^*}$ and $V^{{AD}^*}$.}
\label{GammaVSBudget-Number}
\end{figure}

 Fig.\ref{GammaVSBudget-Number} illustrates the impact of $\gamma$ on $V_k^{{VW}^*}$ and $V^{{AD}^*}$. For $k \in \{1, 2\}$, $V_1^{{VW}^*}$ and $V_2^{{VW}^*}$ is increasing in $\gamma$ since $V_k^{VW} =V_k^{M} +V_k^{A}$. Meanwhile, since $\phi_1$ and $\phi_2$ increase in $\gamma$, and the impact of $\sum_{k\in\bar{\mathcal{K}}} \omega_k G \phi_k N {\chi}_k^N(p_k^*)\bar{ \tau}_k(P_a, p_k^*,M_k^*)$ in (\ref{ADer_Utility_Prob}) is dominant, thus, $V^{{AD}^*}$ increases in $\gamma$. For $k \in \{9, 10\}$, $V_9^{{VW}^*}$, $V_{10}^{{VW}^*}$ and $V^{{AD}^*}$ are decreasing in $\gamma$, as shown in the dashed lines.

 In addition, we can conclude from Fig.\ref{GammaVSBudget-Number} that the impact of $\gamma$ on $V_k^{{VW}^*}$ and $V^{{AD}^*}$ depends on the popularity of the websites that participate the AD spaces game: if the websites are with high-ranking popularity, both $V_k^{{VW}^*}$ and $V^{{AD}^*}$ are increasing in $\gamma$; else, both $V_k^{{VW}^*}$ and $V^{{AD}^*}$ are decreasing in $\gamma$.

 \subsection{The impact of AD space maintenance cost: $C_k$}
In this section, we will verify the impacts of $C_k$ on each key indicator and utility.
We first let $\sigma_k = 1.25\sigma_k^T$, $\gamma = 0.5$ and $C_1 = \cdots = C_{10} \in [1,2, \cdots, 10]$.
The remaining parameter settings are the same as section \ref{sigma}.

\begin{figure}[htbp]
\centering
\includegraphics[width=0.5\textwidth]{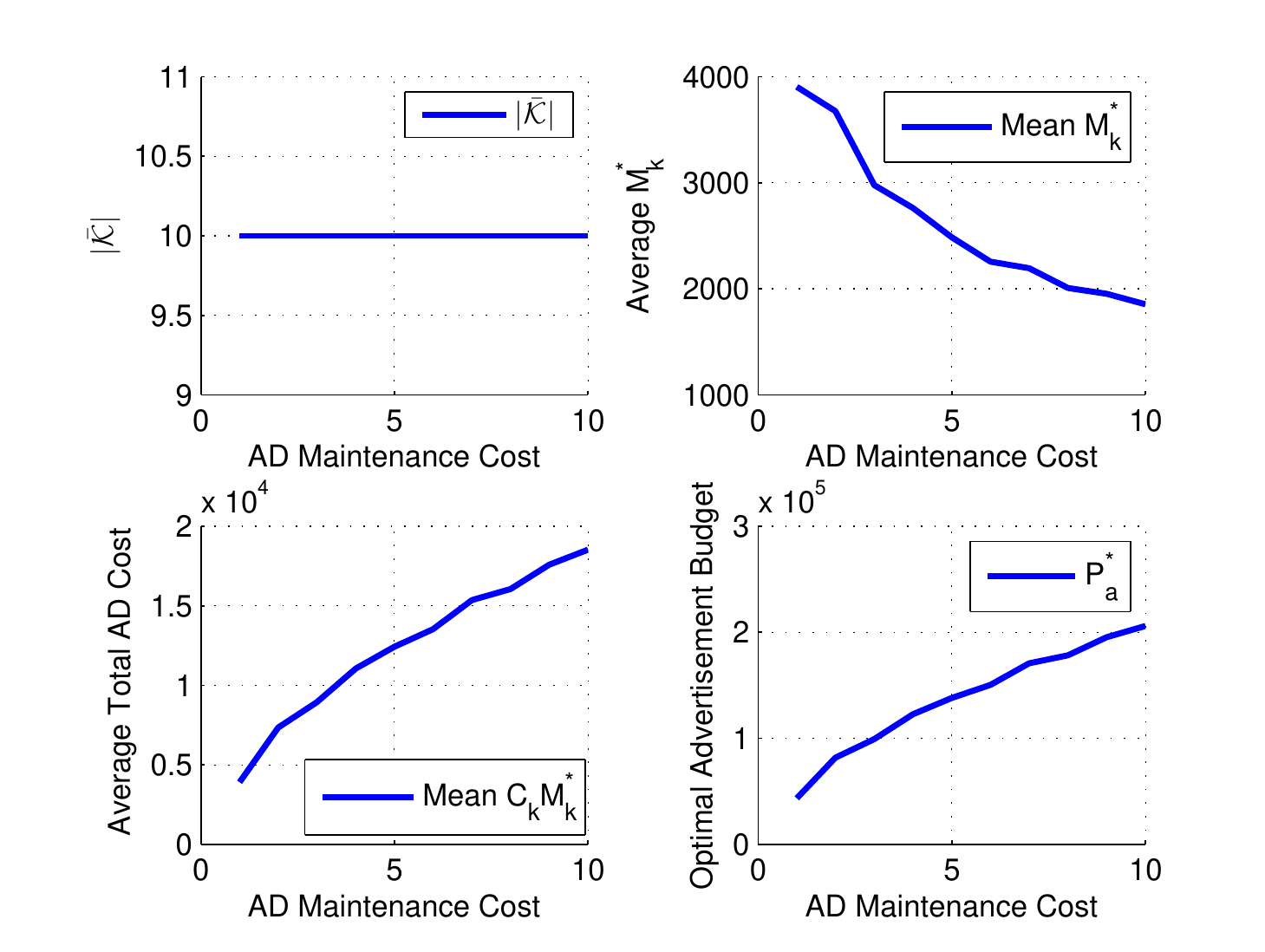}
\caption{The impacts of different values of $C_k$ on $|\bar{\mathcal{K}}|$, $\text{Mean}~M_k^*$, $\text{Mean}~C_kM_k^*$ and $P_a^*$.}
\label{CostVSParameter}
\end{figure}

Fig.\ref{CostVSParameter} presents the impact of $C_k$ on $|\bar{\mathcal{K}}|$, $\text{mean}~M_k^*$, $\text{mean}~C_kM_k^*$ and $P_a^*$. Firstly, since all the AD space maintenance costs are equal in the setting, all the websites satisfy the conditions in Proposition \ref{C_Range} and Proposition \ref{K_Structure}, all of them participate in the advertising budget game, i.e., $|\bar{\mathcal{K}}|=10$, as shown in the first sub-fig in Fig. \ref{CostVSParameter}. Then, we can also observe that $M_k^*$ decreases in $C_k$. This is because when $C_k$ increases, $A_k = {\frac{(|\bar{\mathcal{K}}|-1)}{\sum_{j \in \bar{\mathcal{K}}} C_j}}\left( 1- {\frac{(|\bar{\mathcal{K}}|-1)C_k }{\sum_{j \in \bar{\mathcal{K}}} C_j}}\right)$ decreases, we know that $\exp\left(-{\frac{A_kP_{a}^*}{B_k}}\right)$ increases by (\ref{first_order}), thus $M_k^* = P_a^*A_k$ decreases. Next, since the growth rate of $C_k$, i.e., 9, is larger than the decrement rate of $\text{mean}~M_k^*$, i.e., -0.5, therefore, the average total AD cost $C_kM_k^*$ is increasing in $C_k$. Lastly, we find that $P_a^*$ is increasing in $C_k$. This is due to the fact that the decrement rate of mean $M_k^*$, i.e., -0.5, is less than that of $A_k $, i.e., -0.9, and the fact $M_k^* = P_a^*A_k$, thus, $P_a^*$ must be an increasing function of $C_k$.

\begin{figure}[htbp]
\centering
\includegraphics[width=0.5\textwidth]{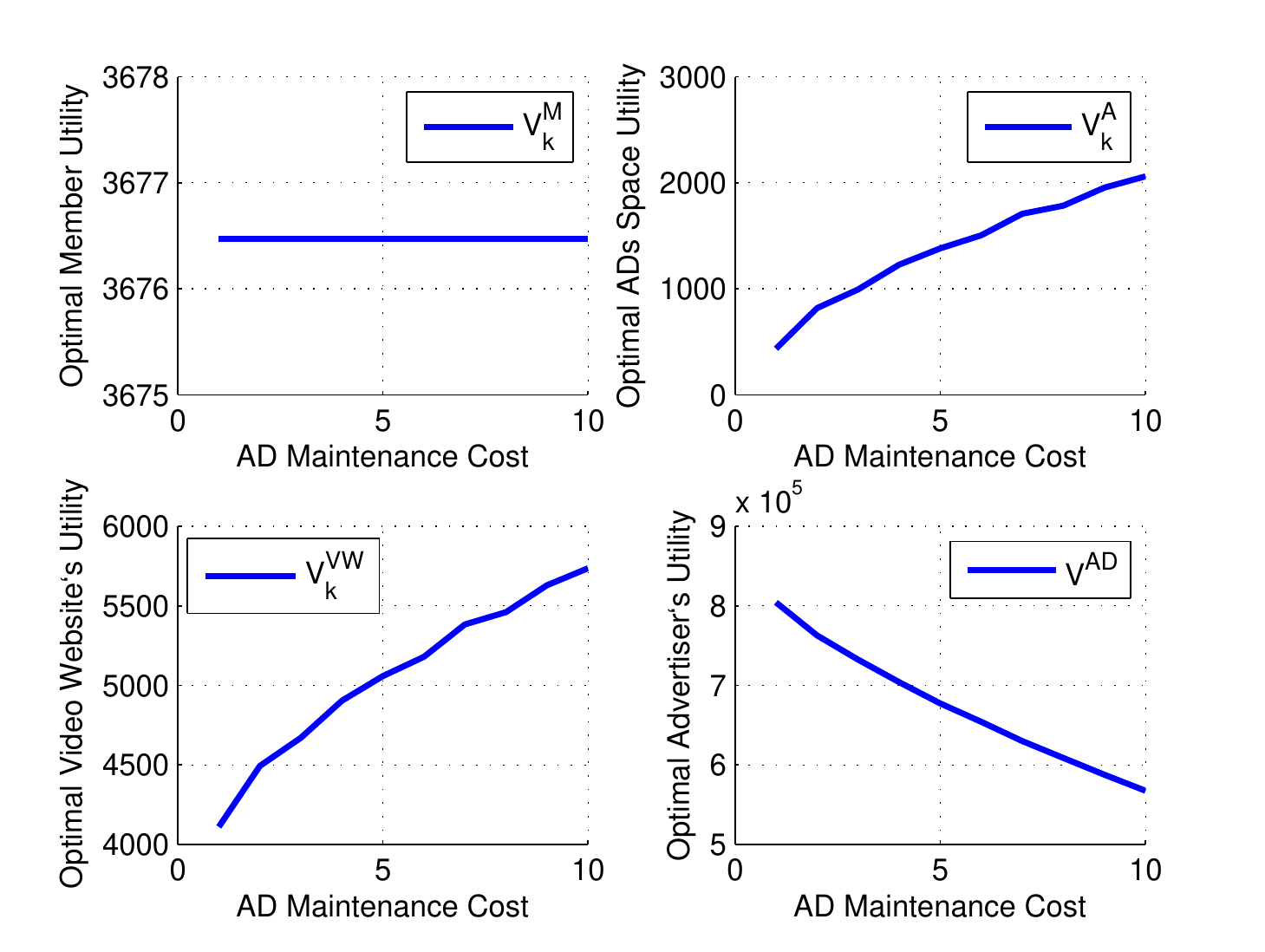}
\caption{The impacts of different values of $C_k$ on each utility.}
\label{CostVSUtility}
\end{figure}

Fig.\ref{CostVSUtility} shows the impact of $C_k$ on each utility. First, since the member utility $V_k^M$ is independent of $C_k$, $V_k^{M^*}$ keeps the same when $C_k$ varies from 1 to 10.
In addition, since $P_a^*$ is increasing in $C_k$ from Fig.\ref{CostVSParameter}, by the derived formula $V_k^{A^*} = P_a^*\left( 1- {\frac{(|\bar{\mathcal{K}}|-1)C_k}{\sum_{j \in \bar{\mathcal{K}}} C_j}}\right)^2$, we know that $V_k^{A^*}$ is increasing in $C_k$. Furthermore, we can  see that $V_k^{{VW}^*}$ is increasing in $C_k$ according to the formula $V_k^{VW} =V_k^{M} +V_k^{A}$ and the monotonicity of $V_k^{M^*}$ and $V_k^{A^*}$. Lastly, $V_k^{{AD}^*}$ is decreasing in $C_k$. This is because with the increase in $C_k$, the increase in $\text{mean}~M_k^*$ (from Fig.\ref{CostVSParameter}) leads to the decrease in $\tau_k(P_a, p_k^*,M_k^*) $, and further leads to the decrease in $\sum_{k\in\bar{\mathcal{K}}} \omega_k G \phi_k N {\chi}_k^N(p_k^*)\bar{ \tau}_k(P_a, p_k^*,M_k^*)$ in (\ref{ADer_Utility_Prob}). Meanwhile, we know that $P_a^*$ is increasing in $C_k$ from Fig.\ref{CostVSParameter}. Therefore, $V_k^{{AD}^*}$ is decreasing in $C_k$.

\begin{figure}[htbp]
\centering
\includegraphics[width=0.5\textwidth]{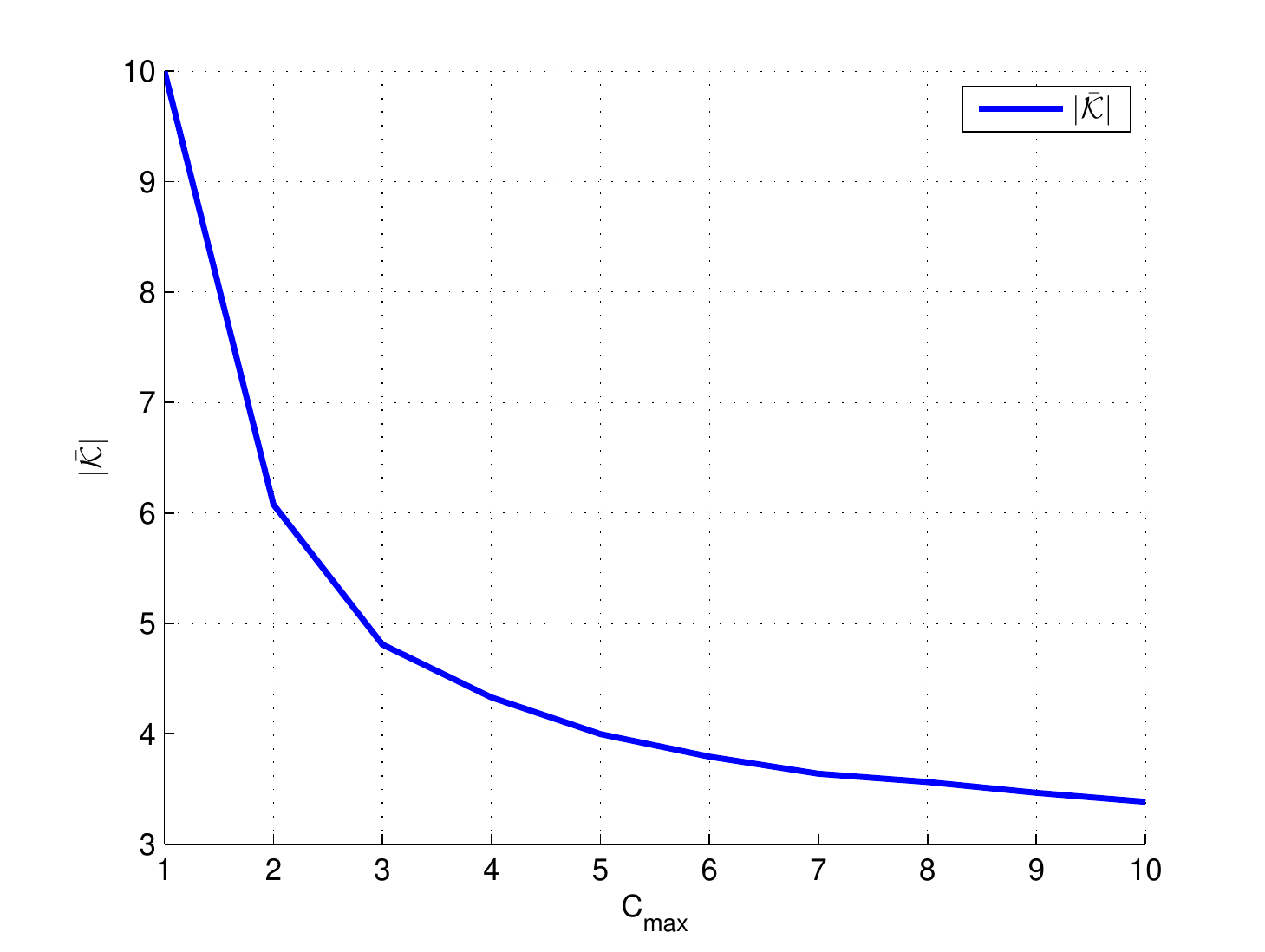}
\caption{The impacts of different values of $C_{\max}$ on $|\bar{\mathcal{K}}|$.}
\label{CostVSKBAR}
\end{figure}

Next, we let $C_k$ follow the uniform distribution over $[1,C_{\max}]$, where $C_{\max}$ varies from 1 to 10 with unit increment.
The other parameter settings are kept the same.
Fig. \ref{CostVSKBAR} illustrates the impact of $C_{\max}$ on the number of video websites participating in the advertising budget game, i.e., $|\bar{\mathcal{K}}|$.
It can be easily observed that $|\bar{\mathcal{K}}|$ decreases in $C_{\max}$.
This is because when $C_{\max}$ increases, the AD costs of all video website become diverse, which increases the possibility of violating the conditions in Proposition \ref{C_Range} and Proposition \ref{K_Structure}.

\subsection{The impact of visiting frequency $\xi$}
\begin{figure}[htbp]
\centering
\includegraphics[width=0.5\textwidth]{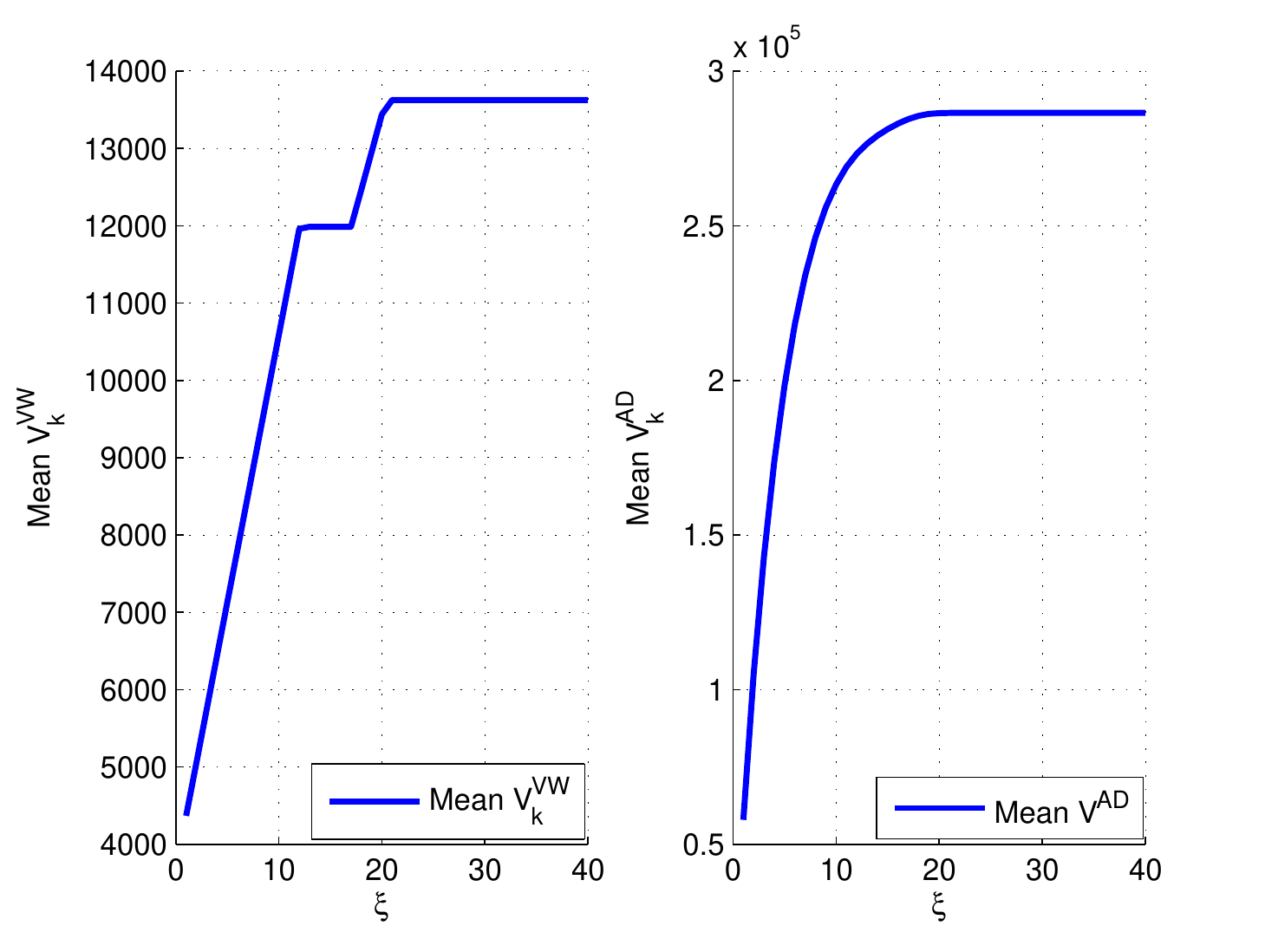}
\caption{The impact of visiting frequency $\xi$ on $V_k^{{VW}^*}$ and $V^{{AD}^*}$.}
\label{XiVSUtility}
\end{figure}

Fig.\ref{XiVSUtility} illustrates the impact of visiting frequency $\xi$ on the utilities of video website and advertiser, i.e., $V_k^{{VW}^*}$ and $V^{{AD}^*}$, respectively.
We choose $\sigma_k = 1.25\sigma_k^T$, $\gamma = 0.5$, $C_k = k$ and let $\xi$ vary from 1 to 40 with increment of 1.
The remaining parameter settings are the same as section \ref{sigma}. We can see that both $V_k^{{VW}^*}$ and $V^{{AD}^*}$ increase in $\xi$ when $\xi \in [0,20]$, and keep the same when $\xi \in [21, 40]$. The reason is that when $\xi \in [0,20]$, There exists at least one AD spaces saturated website, in this case, $P_a^*$ is an increasing function of $\xi$, thus, $V_k^{{VW}^*}$ is increasing in $\xi$ due to $V_k^{VW} =V_k^{M} +V_k^{A}$, where $V_k^{A^*}$ is increasing in $P_a^*$ and $V_k^{M^*}$ is a constant. Meanwhile, $V^{{AD}^*}$ increases in $P_a^*$ in this case, and thus increases in $\xi$. When $\xi \in [21,40]$, there does not exist any AD spaces saturated website, in this case, both $V_k^{{VW}^*}$ and $V^{{AD}^*}$ are independent of $\xi$, therefore, both of them are keep the same even  increasing the value of  $\xi$.

\section{Conclusions}
In this paper, we considered a novel economics model of video websites via \textit{Membership-Advertising Mode} in wireless network, and analyzed the interactions among the advertiser, mobile users and video websites through a three-stage Stackelberg game.
In addition to study the equilibrium of each sub-game, we also investigated the impacts of different \textit{VIP-Member} price coefficient values on the mobile users's watching strategies.
We observed that the optimal subset of video websites participating in the advertising budget game only depended on the advertising space maintenance cost  rather than the advertising budget, and further find out the subset through exploiting the special structure of those cost.
Furthermore, the numerical results revealed the following insights:
1) The utility of video website (advertiser) is decreasing in the \textit{VIP-Member} price coefficient;
2) If the video websites that participate in the advertising budget game are with high-ranking popularity, then, the utility of video website (advertiser) is increasing in \textit{concentration level}, else, the utility is decreasing in \textit{concentration level}.
3) The utility of video website (advertiser) is increasing (decreasing) in the advertising space maintenance cost, and the number of video websites  that participate in the advertising budget game is decreasing if the costs become diverse.
4) If there exists AD spaces \textit{saturated} video website (or visiting frequency is small), the utility of video website (advertiser) is increasing in visiting frequency, if there does not exist any AD spaces \textit{saturated} video website (visiting frequency is sufficiently large), the utility of video website (advertiser) is independent of visiting frequency.

\appendices
\section{Proof of Lemma \ref{Threshold_Relationship}} \label{Proof_Threshold_Relationship}
First, we let ${\frac{(\sigma_k-1)p_k}{Q_k^V- Q_k^R}}={\frac{\sigma_kp_k}{Q_k^V- Q_k^N}}$, we have $\sigma_k = {\frac{Q_k^V-Q_k^N}{Q_k^R-Q_k^N}}$.
We then let ${\frac{(\sigma_k-1)p_k}{Q_k^V- Q_k^R}}={\frac{p_k}{Q_k^R- Q_k^N}}$, and we also have $\sigma_k = {\frac{Q_k^V-Q_k^N}{Q_k^R-Q_k^N}}$.
Therefore, when ${\frac{(\sigma_k-1)p_k}{Q_k^V- Q_k^R}} = {\frac{\sigma_kp_k}{Q_k^V- Q_k^N}} = {\frac{p_k}{Q_k^R- Q_k^N}}$, the price coefficient satisfies $\sigma_k = {\frac{Q_k^V-Q_k^N}{Q_k^R-Q_k^N}}$.
Since $Q_k^V>Q_k^R>Q_k^N>0$, then we have
\begin{align}\label{Price_Coefficient_Threshold}
{\frac{Q_k^V-Q_k^N}{Q_k^R-Q_k^N}} &= {\frac{Q_k^V-Q_k^R+Q_k^R-Q_k^N}{Q_k^R-Q_k^N}}
                  = {\frac{Q_k^V-Q_k^R}{Q_k^R-Q_k^N}} +1 >1.
\end{align}
We denote $\sigma_k^T \triangleq {\frac{Q_k^V-Q_k^N}{Q_k^R-Q_k^N}}$ as the \textit{price coefficient threshold} for website $k$.
Note that $\sigma_k \in (1, \infty)$, we know from (\ref{Price_Coefficient_Threshold}) that $\sigma_k^T$ satisfies the requirement of the feasible domain.

We respectively take the first-order derivation of ${\frac{(\sigma_k-1)p_k}{Q_k^V- Q_k^R}}$, ${\frac{\sigma_kp_k}{Q_k^V- Q_k^N}}$ and ${\frac{p_k}{Q_k^R- Q_k^N}}$ with respect to $\sigma_k$, i.e.,
\begin{align}
&{\frac{\partial\left({\frac{(\sigma_k-1)p_k}{Q_k^V- Q_k^R}}\right)}{\partial\sigma_k}} = {\frac{p_k}{Q_k^V- Q_k^R}},\label{derivation1}\\
&{\frac{\partial\left({\frac{\sigma_kp_k}{Q_k^V- Q_k^N}}\right)}{\partial\sigma_k}} = {\frac{p_k}{Q_k^V- Q_k^N}}, \label{derivation2}\\
&{\frac{\partial\left({\frac{p_k}{Q_k^R- Q_k^N}}\right)}{\partial\sigma_k}} = 0. \label{derivation3}
\end{align}
Since $Q_k^V- Q_k^N>Q_k^V- Q_k^R>0$, we have ${\frac{p_k}{Q_k^V- Q_k^R}}>{\frac{p_k}{Q_k^V- Q_k^N}}>0$.
Note that ${\frac{(\sigma_k-1)p_k}{Q_k^V- Q_k^R}} = {\frac{\sigma_kp_k}{Q_k^V- Q_k^N}} = {\frac{p_k}{Q_k^R- Q_k^N}}$ at $\sigma_k=\sigma_k^T$, thus, when $1<\sigma_k \le \sigma_k^T $, we have ${\frac{(\sigma_k-1)p_k}{Q_k^V- Q_k^R}} \le {\frac{\sigma_kp_k}{Q_k^V- Q_k^N}} \le {\frac{p_k}{Q_k^R- Q_k^N}}$;
when $\sigma_k > \sigma_k^T $, we have ${\frac{(\sigma_k-1)p_k}{Q_k^V- Q_k^R}} > {\frac{\sigma_kp_k}{Q_k^V- Q_k^N}} > {\frac{p_k}{Q_k^R- Q_k^N}}$.

The proof of Lemma \ref{Threshold_Relationship} is completed. \hfill $\Box$

\section{Proof of Theorem \ref{Optimal web price}} \label{Proof_Optimal web price}
We respectively take the first-order and second-order derivation of $V_k^M(p_k)$ with respect to $p_k$, i.e.,
\begin{align}
{\frac{\partial V_k^M(p_k)}{\partial p_k}} &=  \sigma_k  - 2p_k {\frac{(\sigma_k-1)^2( Q_k^R- Q_k^N)+( Q_k^V- Q_k^R)}{\theta_{\max}( Q_k^V- Q_k^R)( Q_k^R- Q_k^N)}}\\
{\frac{\partial^2 V_k^M(p_k)}{\partial p_k^2}} &=  - 2{\frac{(\sigma_k-1)^2( Q_k^R- Q_k^N)+( Q_k^V- Q_k^R)}{\theta_{\max}( Q_k^V- Q_k^R)( Q_k^R- Q_k^N)}}
\end{align}
Meanwhile, the constraint $\theta_k^{T_{12}}(p_k) \le \theta_{\max}$ can be rewritten as $p_k \le {\frac{\theta_{\max}( Q_k^V- Q_k^R)}{(\sigma_k-1)}}$.
Since ${\frac{\partial^2 V_k^M(p_k)}{\partial p_k^2}}<0$, there exists unique optimal solution for the problem  (\ref{Problom_utility_webprice}) in the feasible domain $p_k \in [0, {\frac{\theta_{\max}( Q_k^V- Q_k^R)}{(\sigma_k-1)}}]$.
We let ${\frac{\partial V_k^M(p_k)}{\partial p_k}} = 0$, then, we have
\begin{align}
p_k^* = \left[ {\frac{\sigma_k\theta_{\max}( Q_k^V- Q_k^R)( Q_k^R- Q_k^N)}{2\left[ (\sigma_k-1)^2( Q_k^R- Q_k^N)+( Q_k^V- Q_k^R)\right]}}\right]^{\frac{\theta_{\max}( Q_k^V- Q_k^R)}{(\sigma_k-1)}}
\end{align}
for $k \in \mathcal{K}$.

Thus, the proof of Theorem \ref{Optimal web price} is completed. \hfill $\Box$

\section{Proof of Theorem \ref{Optimal AD Space}} \label{Proof_Optimal AD Space}
First, we should note that $\sum_{j\in \bar{\mathcal{K}}}M_j = \sum_{j\in \mathcal{K}}M_j$,
for website $k$, (\ref{KKT1}) can be rewritten as:
 \begin{align}
\frac{\partial\mathcal{L}}{\partial M_k^*}& = {\frac{\sum_{j\in\mathcal{K}\backslash\{k\}}M_j^*}{\left({\sum_{j\in \mathcal{K}}M_j^*}\right)^2}}P_a-C_k +\beta_k^*
={\frac{\sum_{j\in\bar{\mathcal{K}}\backslash\{k\}}M_j^*}{\left({\sum_{j\in \bar{\mathcal{K}}}M_j^*}\right)^2}}P_a-C_k +\beta_k^*.
 \end{align}
 Letting $\frac{\partial\mathcal{L}}{\partial M_k^*}=0$, we have
 \begin{align}
{\frac{\sum_{j\in\bar{\mathcal{K}}\backslash\{k\}}M_j^*}{\left({\sum_{j\in \bar{\mathcal{K}}}M_j^*}\right)^2}}P_a-C_k +\beta_k^*=0.\label{derivation_1order}
 \end{align}
Since $\beta_k^*= 0$ for $k \in \bar{\mathcal{K}}$ by (\ref{KKT2}), we then sum up (\ref{derivation_1order}) over all the element in set $\bar{\mathcal{K}}$, i.e.,
 \begin{align}\label{transform}
&\sum_{j\in\bar{\mathcal{K}}} \left[{\frac{\sum_{j\in\bar{\mathcal{K}}\backslash\{k\}}M_j^*}{\left({\sum_{j\in \bar{\mathcal{K}}}M_j^*}\right)^2}}P_a-C_k\right]
={\frac{\sum_{j\in\bar{\mathcal{K}}} \sum_{j\in\bar{\mathcal{K}}\backslash\{k\}}M_j^*}{\left({\sum_{j\in \bar{\mathcal{K}}}M_j^*}\right)^2}}P_a -\sum_{j\in\bar{\mathcal{K}}}C_k 
= {\frac{(|\bar{\mathcal{K}}|-1)}{\left({\sum_{j\in \bar{\mathcal{K}}}M_j^*}\right)}}P_a -\sum_{j\in\bar{\mathcal{K}}}C_k=0
 \end{align}
 Therefore, from (\ref{transform}), we have
 \begin{align}\label{Sum_all_M_k}
{\sum_{j\in \bar{\mathcal{K}}}M_j^*} = {\frac{(|\bar{\mathcal{K}}|-1)P_a}{\sum_{j\in\bar{\mathcal{K}}}C_j}}.
 \end{align}
 Substituting (\ref{Sum_all_M_k}) into (\ref{derivation_1order}), we have
 \begin{align}\label{Sum_Partial_M_k}
{\sum_{j\in \bar{\mathcal{K}}\backslash\{k\}}M_j^*} = {\frac{(|\bar{\mathcal{K}}|-1)^2P_aC_k}{(\sum_{j\in\bar{\mathcal{K}}}C_j)^2}}.
 \end{align}
 Letting $M_k^*(P_a) = 0$ for $k \notin \bar{\mathcal{K}}$ and substituting (\ref{Sum_Partial_M_k}) into (\ref{M_k}), we have
\begin{align}
M_k^*(P_a) ={\frac{P_a(|\bar{\mathcal{K}}|-1)}{\sum_{j \in \bar{\mathcal{K}}} C_j}}\left( 1- {\frac{(|\bar{\mathcal{K}}|-1)C_k}{\sum_{j \in \bar{\mathcal{K}}} C_j}}\right)
\end{align}
for all $k \in \bar{\mathcal{K}}$.

Therefore, the proof of Theorem \ref{Optimal AD Space} is completed.  \hfill $\Box$

\section{Proof of Proposition \ref{K_Range}}\label{Proof_K_Range}
According to the assumption of this paper, we know that for any video website $k$, $k \in \mathcal{K}$, there always exists $\sum_{j\in\mathcal{K}\backslash\{k\}}M_j>0$, which implies that $|\bar{\mathcal{K}}|\ge 1$ in the optimal solution since there exists at least a $k' (\neq k)$ such that $M_{k'}^*>0$.
We further assume that $|\bar{\mathcal{K}}|= 1$, that is to say, in the optimal solution, $M_{k'}^*>0$ and $M_{j}^*=0$ for $j \in \mathcal{K}\backslash k'$.
For video website $k'$, $\sum_{j \neq k'} M_j =0$, which contradicts the assumption.
Therefore, we have $|\bar{\mathcal{K}}|\ge 2$ in the optimal solution. \hfill $\Box$

\section{Proof of Proposition \ref{C_Range}}\label{Proof_C_Range}
Note that $|\bar{\mathcal{K}}|\ge 2$ in the optimal solution by Proposition \ref{K_Range}, the term ${\frac{P_a(|\bar{\mathcal{K}}|-1)}{\sum_{j \in \bar{\mathcal{K}}} C_j}}$ in (\ref{Opt_M_k_1}) is always positive. Therefore, the inequality $ 1- {\frac{(|\bar{\mathcal{K}}|-1)C_k}{\sum_{j \in \bar{\mathcal{K}}} C_j}}>0$ must be satisfied for $k\in \bar{\mathcal{K}}$, i.e., $C_k$ satisfies $(|\bar{\mathcal{K}}|-1)C_k < \sum_{j\in\bar{\mathcal{K}}}C_j$. \hfill $\Box$

\section{Proof of Proposition \ref{K_Structure}}\label{Proof_K_Structure}
According to Proposition \ref{C_Range}, $(|\bar{\mathcal{K}}|-1)C_k < \sum_{j\in\bar{\mathcal{K}}}C_j$ indicates that $(\max_{j \in \bar{\mathcal{K}}}C_{j})(|\bar{\mathcal{K}}|-1) < \sum_{j\in\bar{\mathcal{K}}}C_j$.
Assume $k \notin \bar{\mathcal{K}}$, then, we have $[(|\bar{\mathcal{K}}|+1)-1]C_k \ge \sum_{j\in\bar{\mathcal{K}}}C_j+C_k$.
However, we should note that $C_k \le \max_{j \in \bar{\mathcal{K}}} C_j$, which implies that:
\begin{align}
(|\bar{\mathcal{K}}|-1)C_k \le (|\bar{\mathcal{K}}|-1)(\max_{j \in \bar{\mathcal{K}}}C_{j})<\sum_{j\in\bar{\mathcal{K}}}C_j
\end{align}
and
\begin{align}\label{ine_C}
(|\bar{\mathcal{K}}|-1)C_k+C_k < \sum_{j\in\bar{\mathcal{K}}}C_j +C_k.
\end{align}
Letting $\bar{\mathcal{K}'}\triangleq \bar{\mathcal{K}}\cup k$, (\ref{ine_C}) can be rewritten as:
\begin{align}\label{ine_C1}
(|\bar{\mathcal{K}'}|-1)C_k < \sum_{j\in\bar{\mathcal{K}'}}C_j.
\end{align}
which contradicts the assumption. \hfill $\Box$

\section{Proof of Lemma \ref{P_a_UpperBound}}\label{Proof P_a_UpperBound}
Since when $P_a \ge P_a^{T,\max}$, the \textit{AD Watching Probability} $\bar{\tau}_k(P_a, p_k^*,M_k^*) =1 - \exp(-{\frac{\xi}{G}})$ for all $k \in \bar{\mathcal{K}}$, thus, the objective function in (\ref{ADer_Utility_Prob}) can be simplified as $\sum_{k\in\bar{\mathcal{K}}} \omega_k G \phi_k N {\chi}_k^N(p_k^*)(1 - \exp(-{\frac{\xi}{G}})) - P_a$.
It is easy to find that the left term $\sum_{k\in\bar{\mathcal{K}}} \omega_k G \phi_k N {\chi}_k^N(p_k^*)(1 - \exp(-{\frac{\xi}{G}}))$ is a constant, and thus the objective function is a decreasing function in $P_a$ when $P_a\ge P_a^{T,\max}$. \hfill $\Box$

\section{Proof of Lemma \ref{Opt_P_a_relationship}}\label{Proof_Opt_P_a_relationship}
we respectively derive the first-order derivative and the second-order derivative of $F(P_a)$ as follows:
\begin{align}
&{\frac{\partial F(P_a)}{\partial P_a}} = \sum_{k \in \bar{\mathcal{K}}\backslash \mathcal{S}} \omega_{\alpha_k} A_{\alpha_k} \exp\left(-{\frac{A_{\alpha_k}P_a}{B_{\alpha_k}}}\right)-1.\label{1oderde}\\
&{\frac{\partial^2 F(P_a)}{\partial P_a^2}} = \sum_{k \in \bar{\mathcal{K}}\backslash \mathcal{S}} -\omega_{\alpha_k} {\frac{A_{\alpha_k}^2}{B_{\alpha_k}}} \exp\left(-{\frac{A_{\alpha_k}P_a}{B_{\alpha_k}}}\right)
\end{align}
where $A_{\alpha_k} = {\frac{(|\bar{\mathcal{K}}|-1)}{\sum_{j \in \bar{\mathcal{K}}} C_{\alpha_j}}}\left( 1- {\frac{(|\bar{\mathcal{K}}|-1)C_{\alpha_k}}{\sum_{j \in \bar{\mathcal{K}}} C_{\alpha_j}}}\right)$ and $B_{\alpha_k} = G\phi_{\alpha_k} N{\chi}_{\alpha_k}^N(p_{\alpha_k}^*)$.
It is obvious that ${\frac{\partial^2 F(P_a)}{\partial P_a^2}}<0$, thus, the objective function $F(P_a)$ is concave and there exists optimal solution in the feasible domain.
For $\mathcal{S}$ and $\mathcal{S}'$, we let the first-order derivative of the objective function equal to zero, i.e.,
\begin{align}
\sum_{k \in \bar{\mathcal{K}}\backslash \mathcal{S}} \omega_{\alpha_k} A_{\alpha_k} \exp\left(-{\frac{A_{\alpha_k}P_{a,\mathcal{S}}^*}{B_{\alpha_k}}}\right)-1 = 0\label{first_order}\\
\sum_{k \in \bar{\mathcal{K}}\backslash \mathcal{S}'} \omega_{\alpha_k} A_{\alpha_k} \exp\left(-{\frac{A_{\alpha_k}P_{a,\mathcal{S}'}^*}{B_{\alpha_k}}}\right)-1 = 0\label{first_order_1}
\end{align}
Since $\mathcal{S} \subseteq \mathcal{S}'\in \bar{\mathcal{S}}$, combining (\ref{first_order}) and (\ref{first_order_1}), we have
\begin{align}
\sum_{k \in \bar{\mathcal{K}}\backslash \mathcal{S}} \omega_{\alpha_k} A_{\alpha_k} \exp\left(-{\frac{A_{\alpha_k}P_{a,\mathcal{S}}^*}{B_{\alpha_k}}}\right)&=
\sum_{k \in \bar{\mathcal{K}}\backslash \mathcal{S}'} \omega_{\alpha_k} A_{\alpha_k} \exp\left(-{\frac{A_{\alpha_k}P_{a,\mathcal{S}}^*}{B_{\alpha_k}}}\right) +
\sum_{k \in \mathcal{S}'-\mathcal{S}} \omega_{\alpha_k} A_{\alpha_k} \exp\left(-{\frac{A_{\alpha_k}P_{a,\mathcal{S}}^*}{B_{\alpha_k}}}\right)\nonumber\\
&=\sum_{k \in \bar{\mathcal{K}}\backslash \mathcal{S}'} \omega_{\alpha_k} A_{\alpha_k} \exp\left(-{\frac{A_{\alpha_k}P_{a,\mathcal{S}'}^*}{B_{\alpha_k}}}\right)
\end{align}
Since $\omega_{\alpha_k}$, $A_{\alpha_k}$, $B_{\alpha_k}$ and $P_{a,\mathcal{S}}^*$ are non-negative, $\sum_{k \in \mathcal{S}'-\mathcal{S}} \omega_{\alpha_k} A_{\alpha_k} \exp\left(-{\frac{A_{\alpha_k}P_{a,\mathcal{S}}^*}{B_{\alpha_k}}}\right)\ge0$, we have
\begin{align}
&\sum_{k \in \bar{\mathcal{K}}\backslash \mathcal{S}'} \omega_{\alpha_k} A_{\alpha_k} \exp\left(-{\frac{A_{\alpha_k}P_{a,\mathcal{S}}^*}{B_{\alpha_k}}}\right) \le
\sum_{k \in \bar{\mathcal{K}}\backslash \mathcal{S}'} \omega_{\alpha_k} A_{\alpha_k} \exp\left(-{\frac{A_{\alpha_k}P_{a,\mathcal{S}'}^*}{B_{\alpha_k}}}\right).
\end{align}
Meanwhile, since $\exp\left(-{\frac{A_{\alpha_k}P_a}{B_{\alpha_k}}}\right)$ is decreasing in $P_a$, we have $P_{a,\mathcal{S}}^*\ge P_{a,\mathcal{S}'}^*$.

The proof of Lemma \ref{Opt_P_a_relationship} is completed. \hfill $\Box$

\section{Proof of Theorem \ref{Algo2Opt}} \label{Proof_Algo2Opt}
First, we consider the case such that $\mathcal{S} = \bar{\mathcal{S}} _0=\emptyset$ and the corresponding optimal solution for problem in (\ref{New_Opt}) i.e., $P_{a,\mathcal{S}= \bar{\mathcal{S}} _0}^*$, satisfies $P_{a,\mathcal{S}= \bar{\mathcal{S}} _0}^* < P_a^{T,\min}$.
In this case, due to the concave nature of $\sum_{k\in\mathcal{S}} \omega_{\alpha_k} G \phi_{\alpha_k} N {\chi}_{\alpha_k}^N(p_{\alpha_k}^*)\left(1 - \exp\left(-{\frac{\xi}{G}}\right)\right)+
\sum_{k \in \bar{\mathcal{K}}\backslash \mathcal{S}} \omega_{\alpha_k} G \phi_{\alpha_k} N {\chi}_{\alpha_k}^N(p_{\alpha_k}^*)\tau_{\alpha_k}(P_a, p_{\alpha_k}^*,M_{\alpha_k}^*) -P_a$, where $\mathcal{S} \in \bar{\mathcal{S}}$.
The objective function $\sum_{k \in \bar{\mathcal{K}}} \omega_{\alpha_k} G \phi_{\alpha_k} N {\chi}_{\alpha_k}^N(p_{\alpha_k}^*)\tau_{\alpha_k}(P_a, p_{\alpha_k}^*,M_{\alpha_k}^*)  - P_a$ can reach its maximum value at $P_{a,\mathcal{S}= \bar{\mathcal{S}} _0}^* \in [0, P_a^{T,\min})$.
Meanwhile, according to lemma \ref{Opt_P_a_relationship}, we have the objective function $P_a^{T,\min}\ge P_{a,\mathcal{S}= \bar{\mathcal{S}} _0}^* \ge P_{a,\mathcal{S}= \bar{\mathcal{S}} _1}^*\ge \cdots \ge P_{a,\mathcal{S}= \bar{\mathcal{S}} _{|\bar{\mathcal{K}}|-1}}^*$, which implies that $\sum_{k\in\mathcal{S}} \omega_{\alpha_k} G \phi_{\alpha_k} N {\chi}_{\alpha_k}^N(p_{\alpha_k}^*)\left(1 - \exp\left(-{\frac{\xi}{G}}\right)\right)+
\sum_{k \in \bar{\mathcal{K}}\backslash \mathcal{S}} \omega_{\alpha_k} G \phi_{\alpha_k} N {\chi}_{\alpha_k}^N(p_{\alpha_k}^*)\tau_{\alpha_k}(P_a, p_{\alpha_k}^*,M_{\alpha_k}^*) -P_a$ in (\ref{New_Opt}) is decreasing in domain $[P_{a,\mathcal{S}}^*,+\infty)$, $\mathcal{S} \in \bar{\mathcal{S}}$, and further implies that $V^{AD}(P_a\ge P_a^{T,\min})$ in (\ref{ADer_Utility_New1}) is decreasing with its feasible domain.
Therefore, each objective function in (\ref{ADer_Utility_New1}), i.e., $V^{AD}(P_a^{T,\alpha_k}  \le P_a < P_a^{T,\alpha_{k+1}} )$, will reach its maximum value at $P_a^{T,\alpha_k}$, where $k \in \{1,\cdots,|\bar{\mathcal{K}}|-1\}$.
Meanwhile, since $V^{AD}(P_a)$ is continuous in $[0,+\infty)$, we have $V^{AD}(P_{a,\mathcal{S}= \bar{\mathcal{S}} _0}^*) \ge V^{AD}(P_a^{T,\alpha_1}) \ge \cdots \ge V^{AD}(P_a^{T,\alpha_{|\bar{\mathcal{K}}|}})$.
Therefore, we have $P_a^* = P_{a,\mathcal{S}= \bar{\mathcal{S}} _0}^*$.

Next, we consider the condition that $P_{a,\mathcal{S}= \bar{\mathcal{S}} _0}^* < P_a^{T,\min}$ for problem in (\ref{New_Opt}) can not be satisfied, namely, $P_{a,\mathcal{S}= \bar{\mathcal{S}} _0}^* \ge P_a^{T,\min}$. In this case, we have $P_a^* \ge  P_a^{T,\min}$ since $V^{AD}(P_a )$ is continuous in $[0, \infty)$, and increasing in $[0, P_a^{T,\min}]$. Specifically, if $k^*$ exists,
$k^*=\argmax_{k\in \{1,\cdots,|\bar{\mathcal{K}}|-1\}}   {P_{a,\mathcal{S}= \bar{\mathcal{S}}_k}^*\ge P_a^{T,\alpha_k}}$ implies that $P_{a,\mathcal{S}= \bar{\mathcal{S}} _0}^* \ge P_{a,\mathcal{S}= \bar{\mathcal{S}} _1}^* \ge \cdots \ge P_{a,\mathcal{S}= \bar{\mathcal{S}}_{k^*}}^* \ge P_a^{T,\alpha_{k^*}} \ge  \cdots \ge P_a^{T,\min}$ and $P_{a,\mathcal{S}= \bar{\mathcal{S}}_{|\bar{\mathcal{K}}|-1}}^* \le \cdots \le P_{a,\mathcal{S}= \bar{\mathcal{S}}_{k^*+1}}^* \le P_a^{T,\alpha_{k^*+1}} \le \cdots \le P_a^{T,\alpha_{|\bar{\mathcal{K}}|-1}}$.
This means that each $V^{AD}(P_a )$ in (\ref{ADer_Utility_New1}) is increasing in its own feasible domain when $P_a \in \left[0, \min\{P_{a,\mathcal{S}= \bar{\mathcal{S}}_{k^*}}^*, P_a^{T,\alpha_{k^*+1}}\}\right]$, and decreasing in its own feasible domain when $P_a \in \left[ \min\{P_{a,\mathcal{S}= \bar{\mathcal{S}}_{k^*}}^*, P_a^{T,\alpha_{k^*+1}}\},P_a^{T,\max}\right]$ due to the concave nature of function $\sum_{k\in\mathcal{S}} \omega_{\alpha_k} G \phi_{\alpha_k} N {\chi}_{\alpha_k}^N(p_{\alpha_k}^*)\left(1 - \exp\left(-{\frac{\xi}{G}}\right)\right)+
\sum_{k \in \bar{\mathcal{K}}\backslash \mathcal{S}} \omega_{\alpha_k} G \phi_{\alpha_k} N {\chi}_{\alpha_k}^N(p_{\alpha_k}^*)\tau_{\alpha_k}(P_a, p_{\alpha_k}^*,M_{\alpha_k}^*) -P_a$, where $\mathcal{S} \in \bar{\mathcal{S}}$.
Similarly, since $V^{AD}(P_a )$ is continuous in $[0, \infty)$, $k^*=\argmax_{k\in \{0,\cdots,|\bar{\mathcal{K}}|-1\}}   {P_{a,\mathcal{S}= \bar{\mathcal{S}}_k}^*\ge P_a^{T,\alpha_k}}$ can guarantee that $V^{AD}(P_a )$ is increasing in $ \left[0, \min\{P_{a,\mathcal{S}= \bar{\mathcal{S}}_{k^*}}^*, P_a^{T,\alpha_{k^*+1}}\}\right]$, and decreasing in $\left[ \min\{P_{a,\mathcal{S}= \bar{\mathcal{S}}_{k^*}}^*, P_a^{T,\alpha_{k^*+1}}\},P_a^{T,\max}\right]$.
Therefore, $P_a^* = \min\{P_{a,\mathcal{S}= \bar{\mathcal{S}}_{k^*}}^*, P_a^{T,\alpha_{k^*+1}}\}$. If $k^*$ does not exist, through similar analysis, we have $P_a^* = P_a^{T,\min}$.

The proof of Theorem \ref{Algo2Opt} is thus completed.

\bibliographystyle{IEEEtran}
\bibliography{Myreference}

\end{document}